\documentclass[sigconf]{acmart}

\usepackage{amsmath}
\usepackage{algorithm2e}
\usepackage{xspace}
\usepackage{algpseudocode}

\AtBeginDocument{%
  \providecommand\BibTeX{{%
    \normalfont B\kern-0.5em{\scshape i\kern-0.25em b}\kern-0.8em\TeX}}}

\setcopyright{acmcopyright}
\copyrightyear{2022}
\acmYear{2022}
\acmDOI{10.1145/1122445.1122456}
\acmConference[UIST '22]{UIST '22: ACM Symposium on User Interface Software and Technology}{Oct 29--Nov 2, 2022}{Bend, Oregon, USA}
\acmBooktitle{UIST '22: ACM Symposium on User Interface Software and Technology, Oct 29--Nov 2, 2022, Bend, Oregon, USA}
\acmPrice{15.00}
\acmISBN{978-1-4503-XXXX-X/18/06}

\newcommand{\system}{Mixels}

\begin{document}

\title{
\system{}: Fabricating Interfaces using Programmable Magnetic Pixels
}









\author{Martin Nisser}
\affiliation{%
  \institution{MIT CSAIL}
  \streetaddress{32 Vassar st.}
  \city{Cambridge, MA}
  \country{USA}}
\email{nisser@mit.edu}

\author{Yashaswini Makaram}
\affiliation{%
  \institution{MIT CSAIL}
  \streetaddress{32 Vassar st.}
  \city{Cambridge, MA}
  \country{USA}}
\email{ymakaram@mit.edu}

\author{Lucian Covarrubias}
\affiliation{%
  \institution{MIT CSAIL}
  \streetaddress{32 Vassar st.}
  \city{Cambridge, MA}
  \country{USA}}
\email{lucianc@mit.edu}

\author{Amadou Bah}
\affiliation{%
  \institution{MIT CSAIL}
  \streetaddress{32 Vassar st.}
  \city{Cambridge, MA}
  \country{USA}}
\email{abah@mit.edu}

\author{Faraz Faruqi}
\affiliation{%
  \institution{MIT CSAIL}
  \streetaddress{32 Vassar st.}
  \city{Cambridge, MA}
  \country{USA}}
\email{ffaruqi@mit.edu}

\author{Ryo Suzuki}
\affiliation{%
  \institution{University of Calgary}
  \streetaddress{2500 University Dr NW}
  \city{Calgary, AB}
  \country{Canada}}
\email{ryo.suzuki@ucalgary.ca}

\author{Stefanie Mueller}
\affiliation{%
  \institution{MIT CSAIL}
  \streetaddress{32 Vassar st.}
  \city{Cambridge, MA}
  \country{USA}}
\email{stefanie.mueller@mit.edu}

\renewcommand{\shortauthors}{Nisser et al.}



\begin{teaserfigure}
  \centering
  \includegraphics[width=0.99\columnwidth]{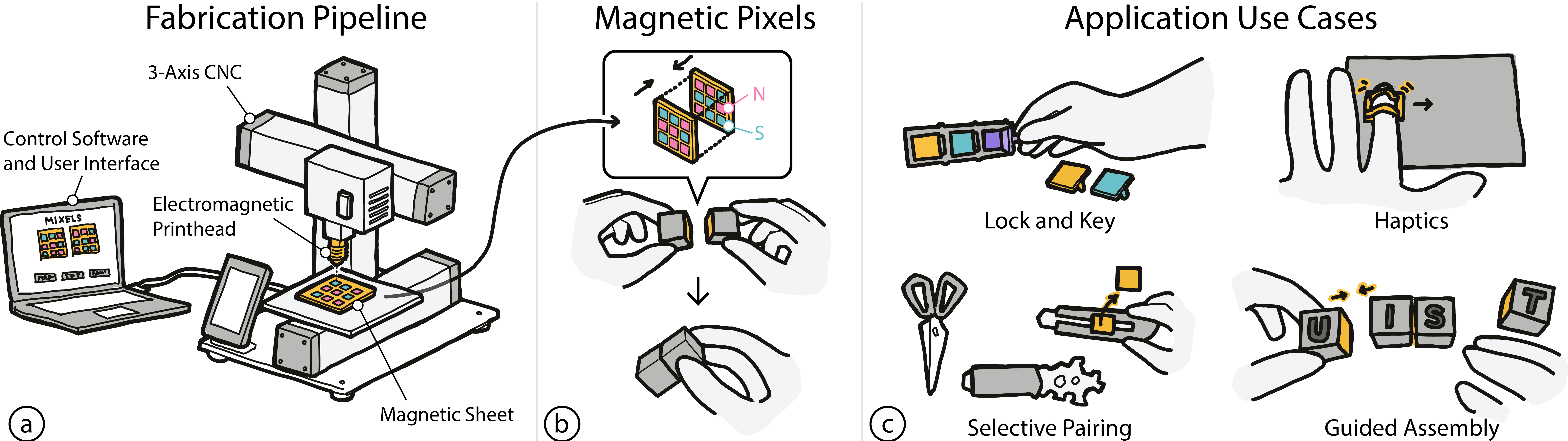}
  \caption{Mixels are interfaces made from programmable magnetic pixels. Users can create Mixels using our custom design and fabrication pipeline: (a)~an inexpensive hardware add-on consisting of an electromagnet and hall effect sensor clipped on to a 3-axis CNC can both write and read magnetic pixels, (b)~the resulting magnetic interfaces can be programmed to attract and repel only in specific configurations and are otherwise agnostic, (c)~this enables a variety of applications, ranging from selectively paired yet visually identical objects, to haptics and guided assembly.}
  \Description{concept}
  \label{fig:concept}
\end{teaserfigure}

\begin{abstract}

In this paper, we present Mixels, programmable magnetic pixels that can be rapidly fabricated using an electromagnetic printhead mounted on an off-the-shelve 3-axis CNC machine. The ability to program magnetic material pixel-wise with varying magnetic force enables Mixels to create new tangible, tactile, and haptic interfaces. To facilitate the creation of interactive objects with Mixels, we provide a user interface that lets users specify the high-level magnetic behavior and that then computes the underlying magnetic pixel assignments and fabrication instructions to program the magnetic surface. Our custom hardware add-on based on an electromagnetic printhead and hall effect sensor clips onto a standard 3-axis CNC machine and can both write and read magnetic pixel values from magnetic material. Our evaluation shows that our system can reliably program and read magnetic pixels of various strengths, that we can predict the behavior of two interacting magnetic surfaces before programming them, that our electromagnet is strong enough to create pixels that utilize the maximum magnetic strength of the material being programmed, and that this material remains magnetized when removed from the magnetic plotter.

\end{abstract}



\maketitle

\section{Introduction}

Advances in digital fabrication tools have enabled users to fabricate objects with a wide range of properties by modifying physical parameters, such as the color ~\cite{programmablefilament2020}, surface texture~\cite{hapticprint2015}, and compliance~\cite{mueller2013laserorigami,nisser2021laserfactory} of objects. More recently, digital fabrication tools have also been used to fabricate objects with other functional properties, such as custom acoustic~\cite{printedspeakers2014} and optical~\cite{lenticular2021} behaviors.

Magnetic materials, however, still remain far behind this digital fabrication revolution~\cite{kim2018printing} despite the fact that magnetic materials hold great promise for interactive applications. For example, researchers used magnets to create novel tangible interfaces~(\textit{MechaMagnets}~\cite{zheng2019mechamagnets}), to produce custom tactile sensations~(\textit{MagneLayer}~\cite{yasu2020magnelayer}), to guide the assembly of 3D objects~(\textit{DynaBlock}~\cite{suzuki2018dynablock}, \textit{ElectroVoxel}~\cite{nisser2022electrovoxel}, \textit{Stochastic Self-assembly}~\cite{nisser2021programmable}), and to create actuated interfaces (\textit{Programmable Polarities}~\cite{nisser2022stochastic}). However, all of these works either use off-the-shelf magnets (\textit{MechaMagnets}~\cite{zheng2019mechamagnets}, \textit{DynaBlock}~\cite{suzuki2018dynablock}) or require manual construction of the customized magnetic materials (\textit{FluxPaper}~\cite{fluxpaper2015}, \textit{MagneLayer}~\cite{yasu2020magnelayer}). 

More recently, researchers also started to automate the fabrication process of custom magnetic materials. For instance, \textit{Magnetic Plotter}~\cite{yasu2017magnetic} is a 2D plotter that stamps custom magnetic patterns onto magnetic sheets using two permanent magnets to provide interaction. Although this method supports programming a variety of patterns, \textit{Polymagnets}~\cite{polymagnets} introduced a commercial product that expedites the programming of magnetic pixels and developed a method to create magnets with unique applications such as non-contact attachment. However, researchers have not yet investigated how to create a design and fabrication pipeline for programmable magnetic pixels that are selectively attractive, repulsive and agnostic between multiple objects. 

In this paper, we introduce \system{}, the first digital design and fabrication pipeline that creates interfaces based on programmable magnetic pixels that exhibit selectively attractive and repulsive behaviors in specific configurations and that otherwise remain agnostic. \system{} fabrication hardware consists of a hardware add-on with an electromagnet for writing magnetic pixels and a hall effect sensor for reading magnetic pixels mounted on a 3-axis CNC machine. \system{} design interface allows users to create magnetic pixel patterns that exhibit desired attractive, repulsive, or agnostic behaviors. On export, the designed magnetic pixel patterns are translated into fabrication instructions, allowing the plotter to program magnetic faces without manual intervention. Our evaluation shows that our system can reliably program magnetic pixels (within 0.3\% accuracy) of different strengths, that we can identify the polarity of a programmed pixel via our magnetic reader, that our chosen electromagnet is strong enough to create pixels of maximum magnetic strength (310 Gauss in either polarity for our chosen magnetic material), and that our magnetic material remains highly magnetized when removed from the magnetic plotter (exhibiting no detectable attenuation once programmed). Finally, our evaluation shows that we can accurately predict the interaction between two magnetically programmed surfaces, which allows us to create interfaces with strong interactions in a chosen configuration while remaining magnetically agnostic otherwise. We illustrate this with a number of applications that enrich interactions of physical objects through connection, sensing, and tactile sensation.

\vspace{5pt}
\noindent In summary, this paper contributes 
\begin{itemize}
\item A hardware add-on that can write and read arbitrary 2D magnetic patterns of continuously variable magnetic strengths.

\item A user interface that lets users design and visualize magnetic patterns with selectively attractive and repulsive behaviors.


\item A technical evaluation of our ability to select, program, and read magnetic pixels from magnetic sheets.

\item Novel applications enabled by our programmable magnetic patterns, such as selective pairing, guided assembly, and vibration and haptics.
\end{itemize}

\section{Related Work}


Many HCI researchers have explored the use of magnets to create novel tangible and haptic interactions since magnets are inexpensive and widely available, and have also embedded them into custom 3D printed objects. To further customize magnetic behavior, researchers have also recently started to create custom programmed magnetic sheets.  

\subsection{Magnets for Tangible and Haptic Interaction}
Many researchers have leveraged magnets to create novel tangible and haptic interactions. One approach focuses on using passive magnets with sensing techniques to create new interaction modalities. For example, \textit{Magnetic Appcessories}~\cite{magneticappcessories2013}, \textit{GaussBits}~\cite{gaussbits2013}, \textit{GaussStones}~\cite{gaussstones2014}, and \textit{Geckos}~\cite{leitner2011geckos} employ a magnetometer to detect the states of physical inputs and positions of magnetic tangible tokens. 
Similarly, \textit{MagGetz}~\cite{maggetz2013}, \textit{MagnID}~\cite{magnid2015}, \textit{GaussBricks}~\cite{gaussbricks2014}, \textit{GaussMarbles}~\cite{gaussmarbles2016}, \textit{Magnetic Ring}~\cite{magneticring2019}, \textit{MagneTrack}~\cite{magnetrack2021} also demonstrate various applications including interactive games and input controls. By combining these sensing approaches with interactive touch displays, these systems allow users to manipulate and interact with digital information via the magnetic tangibles.

Another approach focuses on methods to actuate or otherwise manipulate magnets for dynamic behaviors, using both permanent magnets and electromagnets to create actuated tangible and tactile interfaces. For example, \textit{Actuated Workbench}~\cite{actuatedworkbench2002}, \textit{ShiftIO}~\cite{strasnick2017shiftio}, \textit{FluxMarker}~\cite{suzuki2017fluxmarker}, and \textit{Reactile}~\cite{suzuki2018reactile} move and control passive magnets on a 1D or 2D surface to create actuated tangible interfaces. \textit{ZeroN}~\cite{zeron2011} uses an electromagnetic control system to levitate and manipulate permanently magnetic objects for mid-air tangible interaction. 
Using electromagnets, researchers have explored various applications, such as creating haptic interfaces (\textit{Omni}~\cite{omni2020}, \textit{MAGHair}~\cite{maghair2020}), providing drawing guidance (\textit{dePENd}~\cite{depend2013}), and supporting collaboration (PICO~\cite{10.1145/1240624.1240746}). 
Taking inspiration from these works, we explore how programmable magnetic pixels can expand the range of applications and design space of current magnetic interfaces.

\subsection{Embedding Magnetic Behavior into Objects}

Recently, researchers have also used magnets to add functionality to 3D printed objects. For example, \textit{Programmable Polarities}~\cite{nisser2021programmable}, \textit{MagTics}~\cite{magtics2017}, and \textit{3D Printed Electromagnets}~\cite{electromagneticprinter2016} leverage electromagnets to make the 3D printed objects interactive. However, the use of electromagnets requires power to operate, which decreases portability and ease of fabrication. Researchers have thus also utilized passive magnets to build unpowered, inexpensive interactive devices. For example, \textit{Mechamagnets}~\cite{zheng2019mechamagnets} demonstrate how embedded static magnets in 3D printed parts can deliver various mechanical behavior for physical inputs.
Embedded magnets for 3D printed objects are used to provide haptic feedback (\textit{Magneto-Haptics}~\cite{magnetohaptics2018}, Ogata et al.~\cite{ogata2021computational}), construct objects (\textit{Dynablock}~\cite{suzuki2018dynablock}), and prototype electronics (\textit{Oh, Snap!}~\cite{ohsnap2021}, \textit{LittleBits}~\cite{littlebits2009}).
However, existing works only use discretely sourced, off-the-shelf magnets. In contrast, our work explores how to digitally program magnetic materials with custom patterns to create objects capable of new forms of interaction.

\subsection{Programmed Magnetic Sheet}

To create more custom magnetic behaviors, researchers also investigated how to program magnetic sheets. A programmable magnetic sheet is a soft magnetic sheet that can be programmed into a desired magnetic pattern. HCI researchers have demonstrated the great potential of this approach to create custom  tactile, haptic, and tangible interfaces.
For example, \textit{Magnetic Plotter}~\cite{yasu2017magnetic} explored how to fabricate programmable magnetic sheets that can generate various different tactile sensations for haptic interaction. To program the sheet, \textit{Magnetic Plotter} uses a neodymium magnet that is stronger than the sheet, which allows greater flexibility than just using a non-programmed approach such as \textit{Bump Ahead}~\cite{bumpahead2015}.
\textit{FluxPaper}\cite{fluxpaper2015} also explores a magnetically programmable paper, which allows physical movement and dynamic actuation. 
Beyond a simple pattern, \textit{MagneLayer}~\cite{yasu2020magnelayer} introduces a layered approach, which can create more complex 2D patterns by combining different patterns of the sheet.
By leveraging these capabilities, \textit{Magnetact}~\cite{magnetact2019} and \textit{Magnetact Animals}~\cite{magnetactanimals2021} demonstrate various applications in kinetic toys or interactive haptic interfaces for touchscreen devices. Finally, Polymagnets~\cite{polymagnets} is a commercial product that leverages dedicated machinery to program individual pixels, creating magnets with unique applications such as non-contact attachment and rotational locking between two objects. However, researchers have not yet investigated how to create a design and fabrication pipeline for programmable magnetic pixels that 1) supports interfaces with selectively attractive, repulsive and agnostic behaviors between \textit{multiple} objects, 2) supports digitally reading magnetic pixels to allow for local reprogramming, and 3) is inexpensive and reproducible by users; since Polymagnets is a commercial product, the fabrication device is not available to researchers and it is unclear how users are supported in creating magnetic pixel layouts of desired behaviors. Our work is thus the first to develop a digital design and fabrication pipeline to support users in creating interfaces based on programmable magnetic pixels that exhibit selectively attractive and repulsive behaviors in specific configurations and that otherwise remain agnostic. 

\begin{figure*}[t]
  \centering
  \includegraphics[width=1.9\columnwidth]{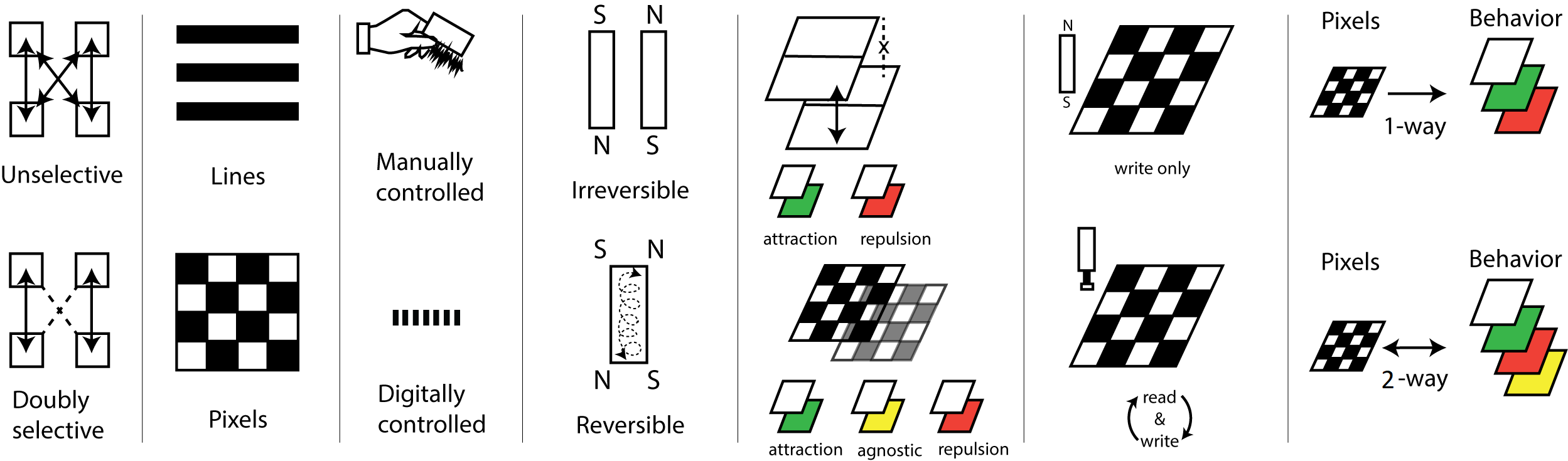}
  \caption{Benefits of \system{}. \system{} leverages its ability to program 2D patterns to construct interfaces that are attractively selective using matrix algebra. It supports designing and plotting arbitrary 2D magnetic patterns in a pixelwise manner, extending the ability to plot lines. It uses digital control by affixing a hardware add-on to a CNC for precision plotting. It uses an electromagnet rather than permanent magnets for plotting, the polarity of which can be digitally controlled. The current through this electromagnet can be continuously regulated to program continuously valued magnetic pixels. By encoding mutually orthogonal matrix values as magnetically programmed pixels, we can program interfaces that are not just attractive or repulsive, but agnostic to each other. Using a Hall effect sensor on our add-on, we can read back programmed magnetic pixels for further manipulation and re-programming. Our user interface supports users to specify behaviors in terms of attraction, repulsion and agnosticism, and can compute the underlying magnet assignments automatically.}
  \Description{Design Space}
  \label{fig:DesignSpace}
\end{figure*}

\section{Mixels: Plotting Magnetic Pixels}

Mixels are programmable magnetic pixels that can be rapidly fabricated using an electromagnetic printhead mounted on an off-the-shelve 3-axis CNC machine. 
To program magnetic pixels, our system consists of (1) the magnetic plotter hardware, (2) the control software that sends commands to the magnetic plotter hardware, and (3) a design user interface that allows the user to specific desired magnetic behaviors. Figure~\ref{fig:concept}  illustrates each of the components of our system and highlights several application scenarios.

\vspace{5pt}
\noindent\textbf{Magnetic Pixels Concept:} At a high-level, the goal of our system is to magnetize each ``pixel'' of a magnetic sheet to create a pixel-wise magnetic pattern. Our hardware applies a magnetic field to specific locations on the magnetic sheet to magnetize and demagnetize pixels one by one, allowing us to program the polarity of each pixel in the soft magnetic sheet. The programmed magnetic pixels allow users to create pairs of magnetic interfaces that (1) attract, (2) repel, and (3) are agnostic (i.e., neither attract nor repel each other) to each other in specific configurations. 
Figure~\ref{fig:DesignSpace} further illustrates the concept and design space of magnetic pixels.
By leveraging this functionality, users can create various applications, ranging from selective pairing, to guided assembly and haptics.

\vspace{5pt}
\noindent\textbf{Design and Fabrication Pipeline:} To support users in creating and reprogramming magnetic pixel patterns, we created a design and fabrication pipeline that automates fabrication and abstracts away the underlying domain knowledge about how magnetic pixels need to be laid out to accomplish attraction, repulsion, and agnosticism via a design tool that allows for high-level input. 
We modified an off-the-shelve 3D printer (SnapMaker) and use its 3-axis motion platform to actuate our magnetic plotter hardware add-on, which consist of an electromagnet and hall effect sensor to both write and read the polarity of magnetic sheets at each pixel individually. The hardware add-on, control software and high-level design input via the user interface allow the user to create uniquely patterned soft magnetic materials. 

\vspace{5pt}
\noindent\textbf{Magnetization and Demagnetization Cycle:} To create magnetic pixels, users insert the magnetic sheet into the plotter. The electromagnet is then turned on and polarized either in 'North' or 'South' direction to increase the magnetic sheet's strength until maximum saturation. The electromagnet is then turned off and the user can removed the magnetic sheet from the magnetic plotter. After removal from the magnetic plotter, the strength of the magnetic sheet drops only marginally, i.e. retains a significant portion of its magnetic strength. If the user wishes to reprogram the magnetic pattern, they can reinsert the magnetic sheet in the magnetic plotter, which demagnetizes the soft magnetic sheet by turning on the electromagnet in the opposite direction of the pixel that needs to be reset, before starting the process of reprogramming the new magnetic pixel value in any polarity (North or South).


\section{Magnetic Plotter Hardware Add-on}

\begin{figure}[b]
  \centering
  \includegraphics[width=0.97\columnwidth]{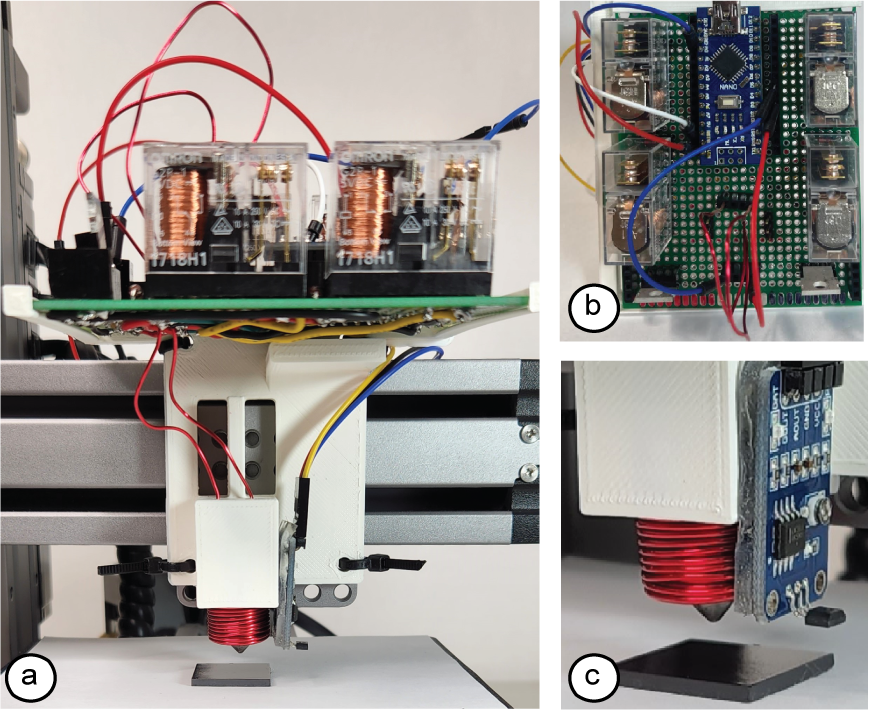}
  \caption{Magnetic plotter: (a)~add-on mounted onto a 3-axis CNC. (b)~add-on electronics viewed from above. (c)~Close-up of the plotting end effector, consisting of an cone-tipped electromagnet for writing and a hall effect sensor for reading magnetic programs.}
  \Description{plotter}
  \label{fig:plotter}
\end{figure}

To make our method available to a wide range of users, we build our magnetic plotter as an inexpensive add-on that can be clipped on to a 3-axis CNC (e.g., we use the SnapMaker 3-in-1). Our add-on consists of an Arduino Nano microcontroller, an electromagnet, an H-bridge, and a hall effect sensor encased in a 3D-printed housing, costing only \$62 in parts. We use our add-on to program commercially available and inexpensive soft magnetic sheet (X-bet, 26 mil thickness) that costs \$0.008/cm$^2$. 

\vspace{5pt}
\noindent\textbf{Magnetic Sheet Material:} We chose a "soft" magnetic sheet material because soft magnetic materials are more easily reprogrammable than hard magnetic materials. This is because soft magnetic materials are more easily demagnetized under a magnetic field. In addition, once demagnetized, they can be reprogrammed (that is, remagnetized) using a weak magnetic field. However, the drawback is that soft magnetic materials have a lower overall magnetic strength than magnetically hard materials. 

\vspace{5pt}
\noindent\textbf{Electromagnet:} We chose an electromagnet over a permanent magnet because we can change the polarity of an electromagnet digitally by changing the direction of current applied to it. In addition, electromagnets allow us to continuously vary the magnetic strength of each pixel by regulating the magnitude of current through them~\cite{yasu2020magnelayer,yasu2017magnetic}. However, cylindrical electromagnets of the same diameter as permanent magnets exhibit less magnetic strength and thus, the resulting magnetic pixels are weaker. To create magnetic pixels of the same strength without trading-off resolution, we therefore shaped the electromagnet into a cone that narrows where it touches the magnetic sheet material. Since the narrow tip concentrates the magnetic flux, our shaped electromagnet creates a stronger magnetic pixel than a permanent magnet for the same pixel size.

\vspace{5pt}
\noindent\textbf{Writing Magnetic Pixels:} To create magnetic pixels, we use an electromagnet that is comprised of a cylindrical permalloy core (10mm diameter, 20mm length) wrapped with 250 turns of 20 AWG wire, with the last 5mm of one end filed to a cone whose tip writes 3mm wide pixels. The core has a relative permeability of $\mu_r=90000$, a factor of 40 greater than most brittle ferrite cores typically used for electromagnets. The electromagnet is coupled to the hardware add-on via a spring-loaded pogo pin, giving it compliance as the plotter touches it to the sheet surface. To drive the electromagnet bidirectionally, we connect it to a full H bridge we built using four relays (Omron 1718H1) driven by 2 MOSFETs (IRFZ44N) and shunted with flyback diodes (1N4752A). Given these features, our method allows us to program magnetic pixels in both polarizations using a single magnet without user intervention. Each pixel requires 0.7 seconds to program, drawing 130W from an offboard power supply to energize the electromagnet during this period and 200mW otherwise. We recorded no excessive heating of the electromagnet, and no observable wear on the CNC even after prolonged use (i.e., we programmed 1500 pixels consecutively to test the add-on's durability in operation). 

\vspace{5pt}
\noindent\textbf{Reading Magnetic Pixels:} To read magnetic pixels programmed onto the sheet, we use a hall effect sensor (MUZHI 49E) conditioned by a voltage comparator (LM393) on a breakout board. This allows us to read both the direction and magnitude of the magnetic polarities of individual pixels and to store these values digitally, which can be later used with our user interface to copy, edit and "paste" (program) pixels, even if the pixel value was previously unknown to the user. In contrast, passive magnetic viewing film can only detect the magnetic strength but cannot distinguish "North" from "South". In addition, since it is an analog method the results seen under viewing film cannot be easily transferred to digital tools.

\begin{figure*}[h]
  \centering
  \includegraphics[width=1.99\columnwidth]{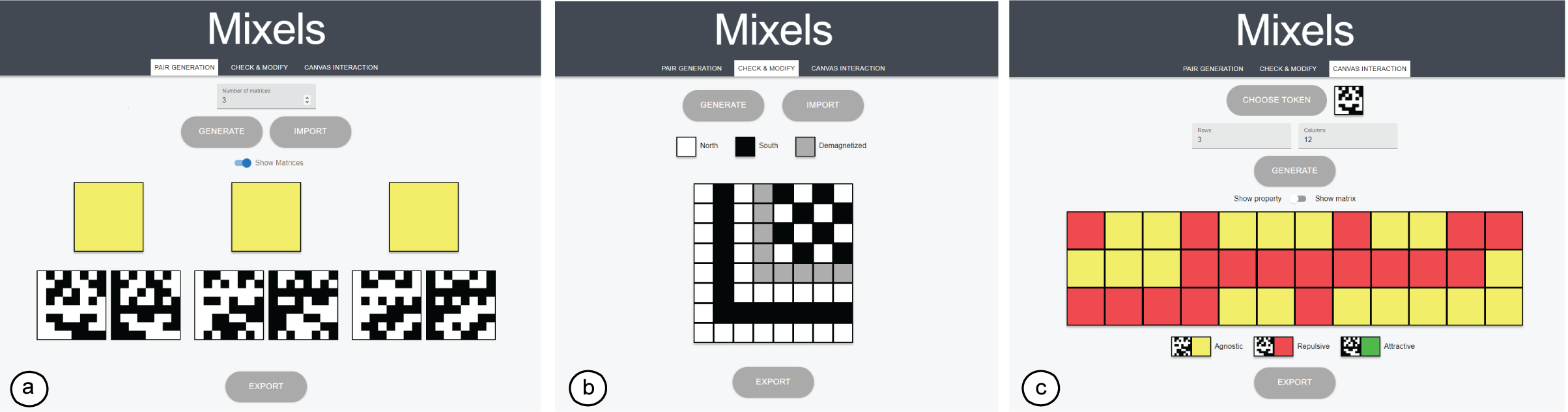}
  \caption{Mixel's user interface with three different design modes: (a)~Individual Pixels: users can edit each pixel directly and assign a value of either 'North', 'South', or 'Demagnetized', (b)~Pair Generation: users can create pairs of matrices which are maximally attractive to each other in specific configurations, but agnostic to other pairs, (c)~Canvas interaction: allows users to define how a smaller matrix should interact with a larger matrix, i.e. either attract, repel, or be agnostic at each location.}
  \Description{UI}
  \label{fig:UI}
\end{figure*}

\section{Control Software}

Our control software sends (1) movement commands to the CNC machine to move the hardware add-on over a specific magnetic pixel, and (2) commands to the electromagnet and hall effect sensor for writing and reading magnetic pixel values at the specific location. The CNC and the hardware add-on's microcontroller (Arduino) are both connected to a laptop via USB cables. We run a local server on the laptop to communicate with the CNC machine and the add-on. A python script on the server accesses the CNC's and hardware add-on's serial channels using the Pyserial library to both send commands and retrieve data via the serial port. 

\vspace{5pt}
\noindent{\textbf{Writing Magnetic Pixels}}: To plot a pattern, users run a python script with the 'plot' parameter and input a previously designed magnetic pixel matrix (.pkl file format). The .pkl file contains the magnetic pixel matrix as a 2D array with elements stored as 1 for 'North' pixels and -1 for 'South' pixels. To plot the matrix, the CNC moves to the first pixel and after the pixel has been programmed by the electromagnet moves translationally (in X or Y) one pixel width at a time. Between each pixel, the CNC rises and descends by 3 mm (in Z) to clear the sheet surface. Unchanged pixels are defined as 0 in the .pkl file and the CNC skips these to save plotting time. After the CNC moved to a specific pixel location, the python script sends a command to the microcontroller connected to the electromagnet to polarize the electromagnet in the correct direction (i.e., either 'North' or 'South') before turning it off. 

\vspace{5pt}
\noindent{\textbf{Reading Magnetic Pixels}}: To read a pattern, users run the python script with the 'scan' parameter and the size of the matrix as input. To read the matrix, the CNC moves to the first pixel and after the pixel has been read by the hall effect sensor moves translationally (in X or Y) one pixel width at a time. Similar to plotting, between each pixel, the CNC rises and descends by 3 mm (in Z) to clear the sheet surface. After the CNC moved to a specific pixel location, the python script sends a command to the microcontroller connected to the hall effect sensor to read the magnetic pixel value. The magnetic pixel readings are then saved as a .pkl file. The .pkl file can then be uploaded to the user interface, which then shows the magnetic polarity at each pixel.

\section{Designing Magnetic Pixel Interactions}

We developed a user interface (Figure~\ref{fig:UI}) to support the design of magnetic pixel layouts with desired behaviors. The user interface allows users to design and visualize custom magnetic patterns and their interactions and automatically converts these patterns into instructions for plotting them with the hardware add-on. The user interface also visualizes magnetic pixel patterns that were scanned from a previously programmed magnetic surface, which allows users to discover, copy or edit previously programmed patterns for new applications. To showcase these features, our user interface supports three design modes: editing individual pixels, pair generation, and Canvas interaction. The user interface is built as a web application on an Amazon Web Services EC2 instance. We created the backend with the Django framework, and used React to design the user interface itself.   

\subsection{Design Modes for Magnetic Pixel Patterns}

\vspace{5pt}
\noindent\textbf{Editing Individual Magnetic Pixels:} In this mode, the user can edit each magnetic pixel individually (Figure~\ref{fig:UI}a). Users first click the 'Generate' button, which creates a 2D matrix of user-defined size consisting of individual pixels. Users can then click each individual pixel to assign them as 'North', 'South', or 'Demagnetized'. The 'Export' button then saves the pixel assignments as a matrix (.pkl file) with values of 1 for 'North' pixels and -1 for 'South' pixels, which is used by the control software to program the magnetic pattern using the hardware add-on. Instead of starting with an empty matrix, users can also use the 'Import' button to load a previously scanned magnetic pixel pattern. Users can then edit individual pixels of the scanned matrix to re-program the scanned pattern for new applications. When hitting the 'Export' button, the exported matrix generated will only account for changed pixels (i.e., saves all unchanged pixels as '0' values), which saves plotting time when only a few pixels require changes. 

\vspace{5pt}
\noindent\textbf{Pair Generation:} This mode allows creating pairs of magnetic surfaces that are attractive to each other when overlapping in one specific configuration but that are not attractive to each other in other configurations or when overlapping with other programmed surfaces. Users can generate a desired number of pairs by entering the number of matrices and then clicking the 'Generate' button. This will then display the magnetic matrix pairs in the user interface. Figure~\ref{fig:UI}a shows that the matrices in each pair are opposite, i.e. where one pixel is black ('North'), the other is white ('South'), such that they perfectly attract when they are directly superimposed. However, when there is a translation of one or more pixels between the surfaces, they will be agnostic to each other due to the orthogonality in these matrices. Similarly, the programmed pairs are also agnostic to the magnetic matrices in the other generated pairs in all configurations. After generating the pairs, users can use the 'export' button to export the matrices collectively for plotting as a set of .pkl files. 

\vspace{5pt}
\noindent\textbf{Canvas interaction:} While the previous mode allowed users to define pair-wise interactions, the 'Canvas interaction' mode allows users to define how a smaller magnetic matrix should interact across a larger magnetic area, or 'canvas' (Figure~\ref{fig:UI}c). To define the interaction, users first use the 'choose token' button to load the smaller matrix, which they previously designed in the user interface. Next, they define the larger magnetic matrix by specifying the number of rows and columns and then initializing the larger magnetic matrix by clicking the 'generate' button. The generated larger magnetic matrix is then automatically partitioned into 'metapixels' whose size is equal to the smaller magnetic matrix. Users can then click each metapixel to assign whether the smaller magnetic matrix should interact with that region attractively, repulsively, or agnostically. In Figure~\ref{fig:UI}c, the larger magnetic matrix can be designed as topological hills and plateaus on a map; a smaller magnetic matrix affixed to a user's finger will feel a repulsive force when moved over a red region, indicating a hill, but remain agnostic elsewhere. After programming the surface, users click the 'Export' button to export the .pkl files for programming both magnetic surfaces.

\subsection{Selective Attraction and Repulsion} 

Both in the 'pair generation' and the 'Canvas interaction' design mode, our user interface generates patterns that attract or repel only when aligned in specific locations and orientations. In all other configurations, those patterns are agnostic.

\vspace{5pt}
\noindent\textbf{Selective Attraction:} To accomplish selective attraction, we generate one matrix of the pair as a so-called 'Hadamard' matrix, and the second matrix of the pair as the complement of the Hadamard. In a Hadamard matrix, every row and every column are orthogonal. Thus, when the complement of the Hadamard is moved across the original Hadamard matrix, they are agnostic in each configuration except when in perfect alignment, which is when they attract. To initialize a Hadamard matrix with orthogonal rows and columns, we use a recursive generation algorithm~\cite{baltieri2010fast}. 

\vspace{5pt}
\noindent\textbf{Selective Repulsion:} Selective repulsion works in the same way to selective attraction, except that the second matrix is identical to the original Hadamard matrix. Thus, when both matrices are moved across each other, they are agnostic in each configuration except when in perfect alignment, which is when they repel. 

\vspace{5pt}
\noindent{\textbf{Multiple Pair Agnosticism}}: While the method described above works to create selective attraction and repulsion for a single pair of magnetic surfaces, the algorithm needs to be extended when multiple pairs of surfaces need to be selectively attractive/repulsive both within the pair and with other pairs. For instance, in the application example shown in Figure~\ref{fig:blocks}, a letter should only connect to its correct neighbor and not to other letters to form the correct word. To accomplish selective attraction/repulsion across multiple pairs of magnetic surfaces, we create new pairs of Hadamard matrices by permuting the rows of the original matrix generated recursively~\cite{baltieri2010fast}. Once many such pairs have been created, we compute the 2D cross-correlation between these permuted matrices and then choose only those matrices that maximize mutual agnosticism between pairs~\cite{Nisser2022preprintIROS}.

\section{Evaluation}

We evaluated how reliably the electromagnet can program the magnetic sheet, if the electromagnet can fully saturate the sheet to endow it with its greatest possible strength, how long the magnetic sheet remains magnetized when removed from the electromagnet, how accurately we can create pixels with continuous magnetic strength, and how accurately we can read magnetic pixel values. 

\subsection{Reliability of Magnetic Programming}

We first evaluated how reliably the magnetic strength of individual pixels can be programmed, which is necessary to ensure consistent behavior.

\vspace{5pt}
\noindent{\textbf{Procedure:}} We collected data by energizing the electromagnet between 0A and 10A, in increments of 1A, in both North and South directions. For each applied current, we recorded the magnetic field at the conic tip of the electromagnet using a Gaussmeter (AlphaLab GM-2). We repeated the measurements 4 times, and computed the mean and standard deviation at each increment.

\begin{figure}[h]
  \centering
  \includegraphics[width=0.97\columnwidth]{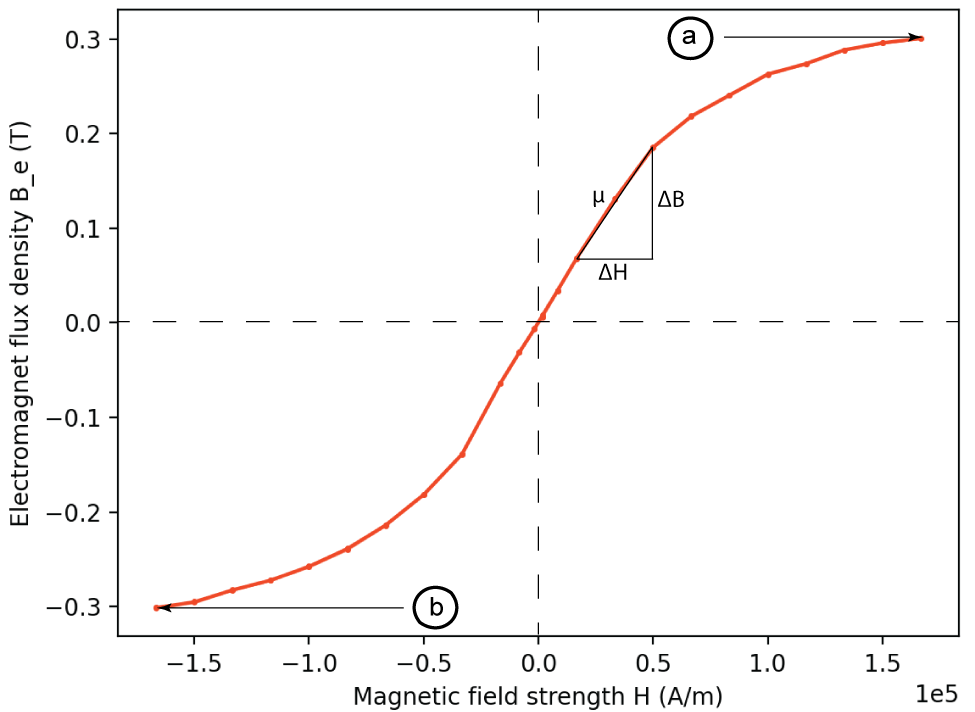}
  \caption{Electromagnet B-H curve. The electromagnet saturates at 0.34T at both (a) positive ('North') and (b) negative ('South') ends.}
  \Description{B-H curve}
  \label{fig:BH-emag}
\end{figure}

\vspace{5pt}
\noindent{\textbf{Results:}} Figure~\ref{fig:BH-emag} shows the resulting magnetization curve, also known as the B-H curve, for the electromagnet. The electromagnet saturates at 0.302T and the measurements exhibit an average standard deviation of 1.01mT, yielding a highly reliable field that can be generated within 0.3\%. The curve is symmetric and exhibits no hysteresis, allowing programming both North and South polarities reliably. The electromagnet can be turned completely off by removing power from the coil; this is illustrated by the curve intersecting the origin, showing that the induced B field collapses when the H field is set to 0, signifying very low coercivity and remanence.

\subsection{Maximum Magnetic Field Strength}

Next, we evaluated if the electromagnet is strong enough to fully saturate the magnetic sheet, which allows creating applications that leverage the material's maximum magnetic strength (flux density).

\begin{figure}[h]
  \centering
  \includegraphics[width=0.97\columnwidth]{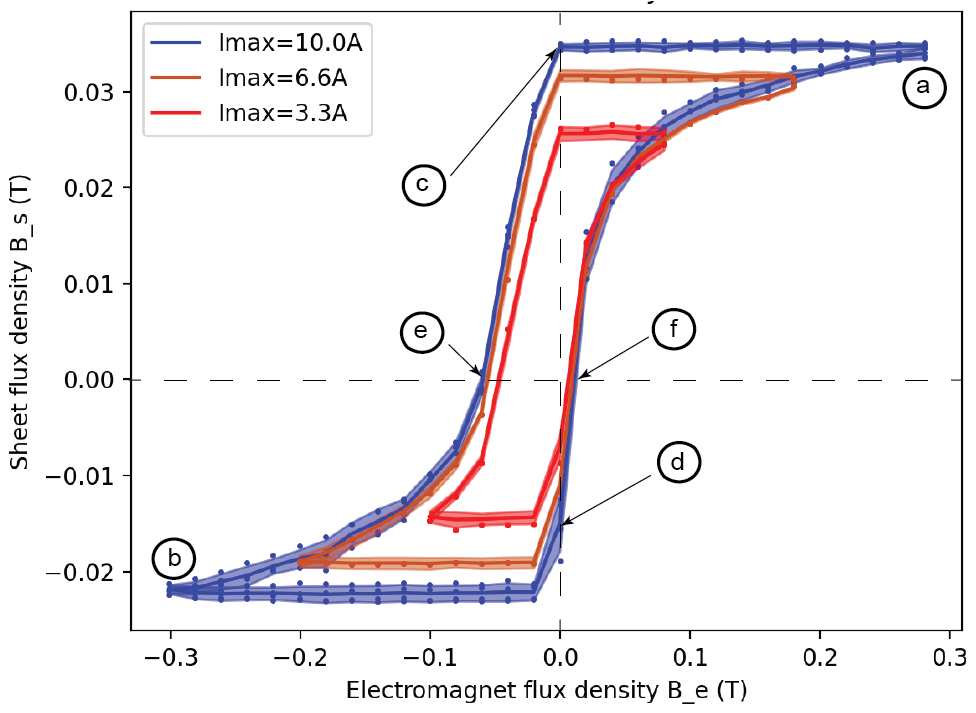}
  \caption{Sheet hysteresis curves when the sheet is fully saturated (major loop, blue) and when it is less than fully saturated (minor loops, red). Major loop is labeled to indicate (a) positive and (b) negative saturation; (c) positive and (d) negative remanence; (e) positive and (f) negative coercivity.}
  \Description{B-H sheet}
  \label{fig:BH-sheet}
\end{figure}

\vspace{5pt}
\noindent{\textbf{Procedure:}} Data was generated by touching the conic end of the electromagnet to the sheet material, energizing the electromagnet, then de-energizing the electromagnet and measuring the field strength of the material sheet where it was programmed. We increased the strength of the electromagnet's magnetic field by increasing the current applied to it from 0A (Figure \ref{fig:BH-sheet}f) upwards in 0.6A increments, measuring the field strength of the sheet where it was programmed with each increment. We continued this procedure until an increase the electromagnet's field produced no additional magnetization of the sheet (Figure \ref{fig:BH-sheet}a).

\vspace{5pt}
\noindent{\textbf{Results:}} Figure \ref{fig:BH-sheet} shows the plotted data starting from \ref{fig:BH-sheet}(f) and ending at \ref{fig:BH-sheet}(a). The procedure showed that an external field generated using $I_{max} = 10A$ was the minimum current required to saturate the sheet. The curve shows that our electromagnet design is strong enough to saturate the sheet (at 0.0344T), generating as much force as possible for applications. In contrast, an equivalently sized pixel programmed by a cylindrical permanent magnet (neodymium, 3mm diameter, 6mm length) was 0.032T.

\subsection{Permanence of Magnetic Field}

We also evaluated the ability of pixels programmed in the magnetic sheet to stay magnetized after programming; their remanence, and the ease with which they can be de-magnetized; their coercivity. 

\vspace{5pt}
\noindent{\textbf{Procedure:}} We begun by following the procedure used to evaluate saturation, detailed above, using $I_{max} = 10A$ to saturate the sheet (Figure \ref{fig:BH-sheet}a). We then de-energized the electromagnet, reducing the current in increments of 0.6A, until the applied current was 0A, and measured the field strength of the material. The magnetic field strength of the sheet with the electromagnet turned off is shown in Figure \ref{fig:BH-sheet}(c), and is the material's remanence. We then continued by applying current to the electromagnet in the opposite direction; a negative current in Figure \ref{fig:BH-sheet}, increasing the magnitude of this negative current in increments of 0.6A until the field strength of the sheet was 0T. This is indicated in Figure \ref{fig:BH-sheet}(e), and is the material's coercivity. The same procedure used to generate the curve \ref{fig:BH-sheet}(f,a,c,e), is used to plot the remaining curve \ref{fig:BH-sheet}(e,b,d,f) by continuing to polarize the electromagnet in the opposite direction starting from \ref{fig:BH-sheet}(e). We repeated this 4 times and computed the mean and standard deviation at each increment.

\vspace{5pt}
\noindent{\textbf{Results:}} Unlike for the electromagnet, the magnetization curve for the magnetic sheet generated in Figure \ref{fig:BH-sheet} reveals a phenomenon known as magnetic hysteresis, which is the dependence of the current magnetization of the sheet on its magnetic history. The hysteresis generated by saturating the material, shown in blue, is known as the major loop. The asymmetry in this hysteresis is what enables the sheet to stay magnetized after programming. The hysteresis curve indicates a low coercivity, indicating that the material is easily re-programmed with a weak external magnetic field, and exhibits high remanence, illustrated by the negligible attenuation of the material's magnetic field strength after the electromagnet is turned off following saturation. The average standard deviation for the major loop was 6.81mT, and the standard deviation at each point is illustrated in the figure to indicate its repeatability.

\subsection{Continuous Magnetic Strength}

We evaluated if we can program pixels with continuous magnetic strengths, by reliably programming the sheet in a way that does not fully saturate it. This results is so-called 'minor' hysteresis loops, as shown by red and orange curves in Figure \ref{fig:BH-sheet}.

\vspace{5pt}
\noindent{\textbf{Procedure:}}  Data used to evaluate the minor curves were gathered using the same strict order as for the major loop, following the loop anti-clockwise. As mentioned previously, an external field generated using $I_{max} = 10A$ fully saturates the sheet. To investigate the effect of not fully saturating the sheet, we therefore chose fields created with lower current using $I_{max} = 3.3A,6.6A$. We then repeated the procedure outlined in section 7.3, beginning at Figure \ref{fig:BH-sheet}(f) and following each loop anti-clockwise back to its starting point. We repeated this 4 times for each $I_{max}$, and computed the mean and standard deviation at each increment. 

\vspace{5pt}
\noindent{\textbf{Results:}} Figure \ref{fig:BH-sheet} shows the two magnetization curves when the field strengths were generated with currents through the electromagnet of 3.3A and 6.6A. The average standard deviation for 3.3A and 6.6A curves are 5.85mT and 6.69mT respectively, and the standard deviation at each point is illustrated in the figure. This shows that the hysteresis loops can be utilized for programming a range of magnetization strengths.

\subsection{Accuracy of Reading Magnetic Pixel Values}

We evaluated how accurately we can read magnetic pixel values with our hall effect sensor.

\vspace{5pt}
\noindent{\textbf{Procedure:}} We programmed 150 North-oriented (with normalized flux of -1) and 150 South-oriented pixels (with normalized flux +1), and recorded the magnetic strength both with the hall effect sensor and with the Gaussmeter to obtain a ground-truth estimate.

\begin{figure}[h]
  \centering
  \includegraphics[width=0.99\columnwidth]{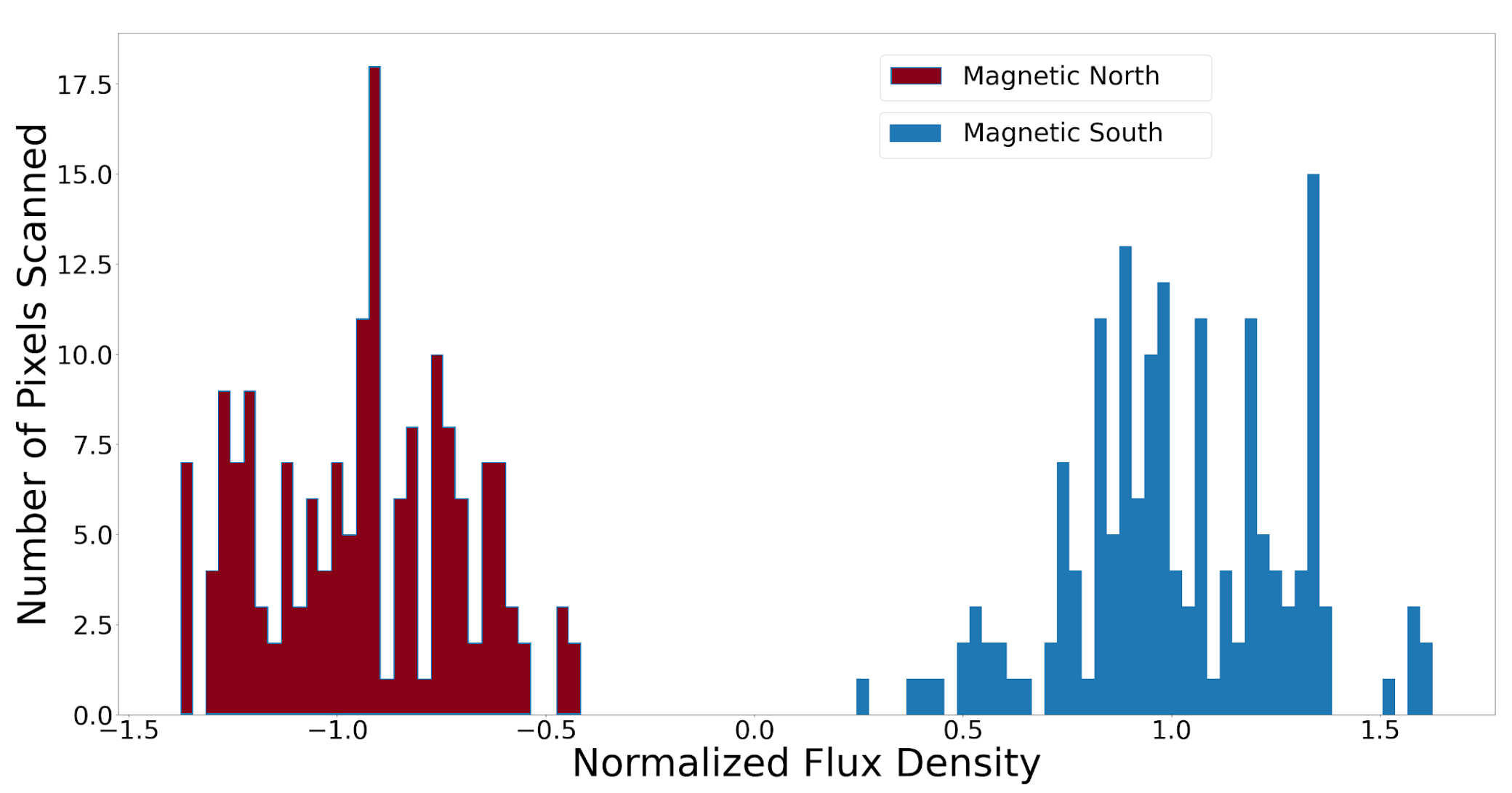}
  \caption{Histogram of magnetic pixel values scanned by hall effect sensor. North- and South-oriented pixels are clearly differentiated.}
  \Description{histogram}
  \label{fig:histo}
\end{figure}

\vspace{5pt}
\noindent{\textbf{Results:}} The recorded magnetic pixel strengths as measured by the hall effect sensor are shown in normalized form as a histogram in Figure~\ref{fig:histo}. North-oriented pixels are shown in red and South-oriented pixels are shown in blue. As can be seen in the figure, the read values exhibit significant noise with recorded values distributed widely around their ground truths of -1 and +1. Ground-truth measurements taken with a Gaussmeter showed that the programmed values were accurate, thus the noise is introduced by the inaccuracy of the hall effect sensor. However, since both the south and north regions have non-overlapping distributions, we can still differentiate between magnetically North- and South-oriented pixels. A higher quality hall effect sensor would be required to accurately measure pixels that were programmed with continuously variable magnetic strengths.

\subsection{Predicted vs Measured Interaction}

Finally, we evaluated how accurately we can predict magnetic interactions in terms of attraction, repulsion and agnosticism.

\begin{figure}[h]
  \centering
  \includegraphics[width=0.99\columnwidth]{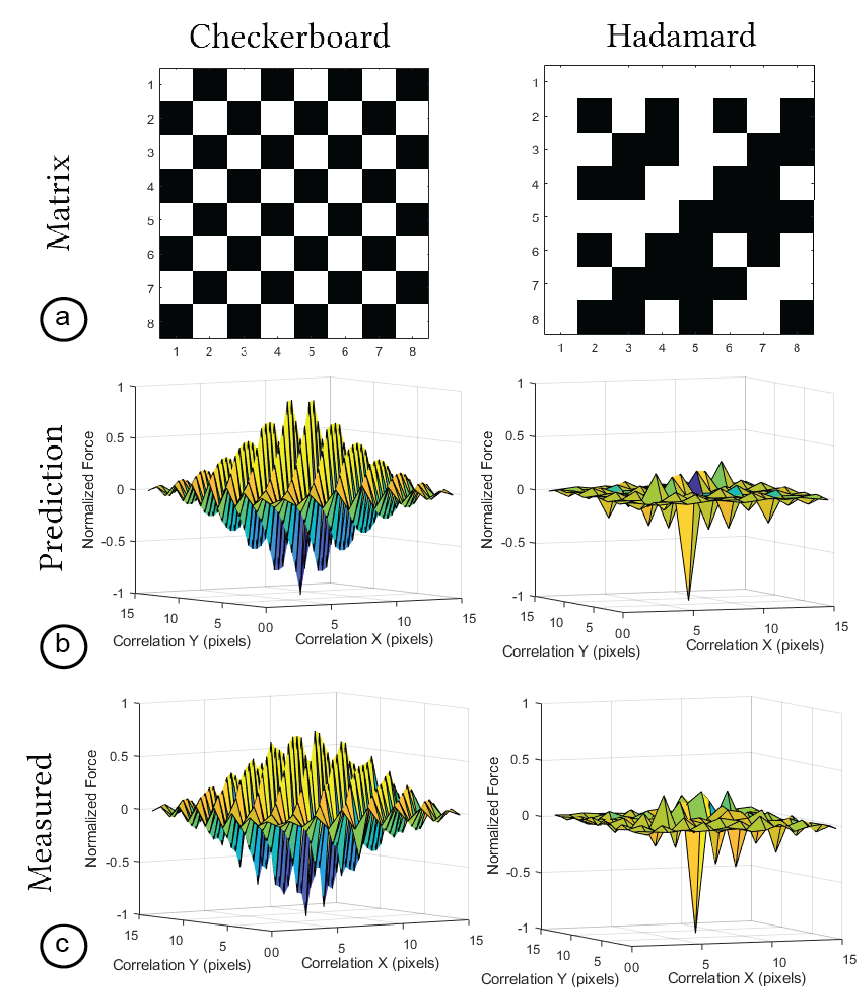}
  \caption{Predicted and measured magnetic interactions when two patterns, a checkerboard and a Hadamard matrix, are cross-correlated with their negatives.}
  \Description{theory-predict}
  \label{fig:theory-predict}
\end{figure}

\vspace{5pt}
\noindent{\textbf{Procedure:}} We created two magnetic pixel designs in our user interface (Figure~\ref{fig:theory-predict}a): (1)~an 8x8 checkerboard pattern of 'North' and 'South' programmed pixels, and (2)~an order-8 Hadamard matrix. We also created their complement matrices: the corresponding magnetic pixel pattern produced by multiplying each matrix by -1. We then translated each matrix pixel-by-pixel, in both X and Y directions, across its complement and evaluated the resulting force at each increment. 

\vspace{5pt}
\noindent{\textbf{Simulation Results:}} Figure~\ref{fig:theory-predict}(b) shows the predicted interaction between each matrix. We generated the predicted values by computing the normalized cross-correlation between each matrix and its complement. This effectively implies taking the sum of all attractive (-1) and repulsive (+1) pixel interactions, and dividing by the number of pixels. As a result, cross-correlation values of -1 designate perfect attraction, +1 is perfect repulsion and 0 is agnosticism. When the checkerboard matrix and Hadamard matrix are centered on their respective complements, they are by definition attractive at every pixel, yielding a value of -1. Elsewhere, the checkerboard produces oscillating attractive and repulsive interactions with every pixel-wise translation, whereas the Hadamard remains perfectly agnostic in pure X-translation and Y-translation, and maximally agnostic for mixed translation. 

\vspace{5pt}
\noindent{\textbf{Physical Results:}} Figure~\ref{fig:theory-predict}(c) shows measured data. We measured this data by first programming the patterns on magnetic sheets (25mm side square). We then affixed one magnetic sheet onto a scale (KUBEI pocket, 0.1mN accuracy) placed on the CNC baseplate. The other sheet was mounted onto the CNC arm, which translated the patterns pixel-wise one pixel at a time (while keeping the magnetic sheet in a planar orientation and 0.5mm apart from the mounted sheet). We then recorded the force at each location using the scale. As can be seen in Figure~\ref{fig:theory-predict}, the measured data corresponds well visually with the simulated cross correlation, showing we can predict magnetic interactions between arbitrarily programmed magnetic sheets accurately before physically programming them. The two square faces of 25mm side length were measured to exhibit an attractive force of 1.09N, corresponding to 1.74 kPa. In shear, the faces could withstand 1.31N, a high value likely caused by the exceptionally high friction coefficient (1.15) of rubber.




\section{Applications}


We built a number of applications to demonstrate how pixel-wise programmed magnetic interfaces enable novel use cases. 

\begin{figure}[h]
  \centering
  \includegraphics[width=0.9\columnwidth]{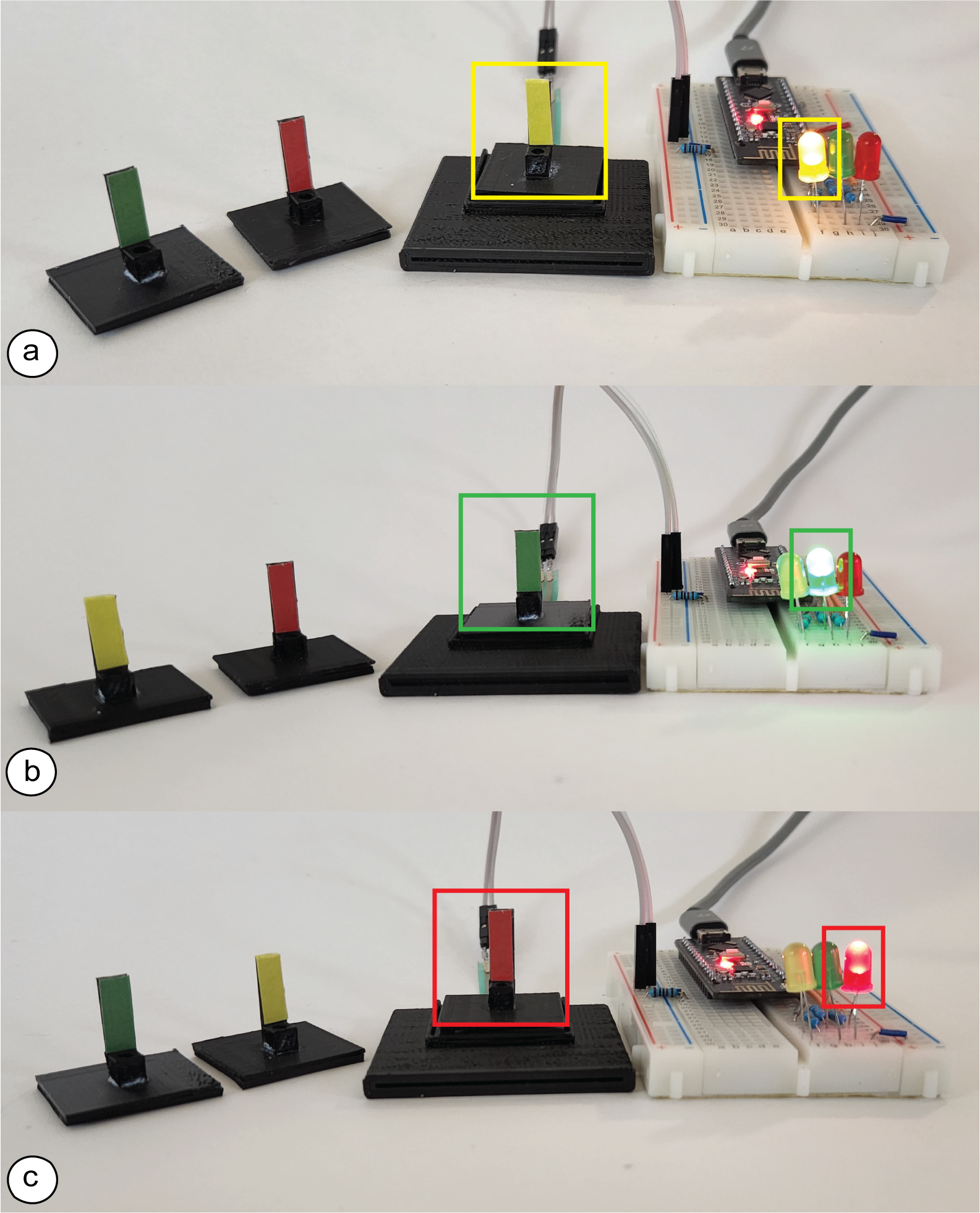}
  \caption{Each magnetic face exhibits a unique force profile when placed on a second magnet, which can be detected via a pressure sensor that then lights up the corresponding LED.}
  \Description{lights}
  \label{fig:lights}
\end{figure}

\vspace{5pt}
\noindent{\textbf{Unique Force Profiles:} Figure~\ref{fig:lights} shows how we can create objects that exhibit unique force profiles. We programmed three magnetic faces (marked green, yellow, and red) with different magnetic patterns and affixed them to the bottom of 3D-printed handles. When each of the handles is placed on a platform housing a second magnet below a pressure sensor, the unique force profile is registered by the microcontroller connected to the pressure sensor and subsequently lights up the associated colored LED. We created the different force profiles by designing the magnetic patterns in the user interface using the mode for directly programming individual pixels.}

\begin{figure}[h]
  \centering
  \includegraphics[width=0.9\columnwidth]{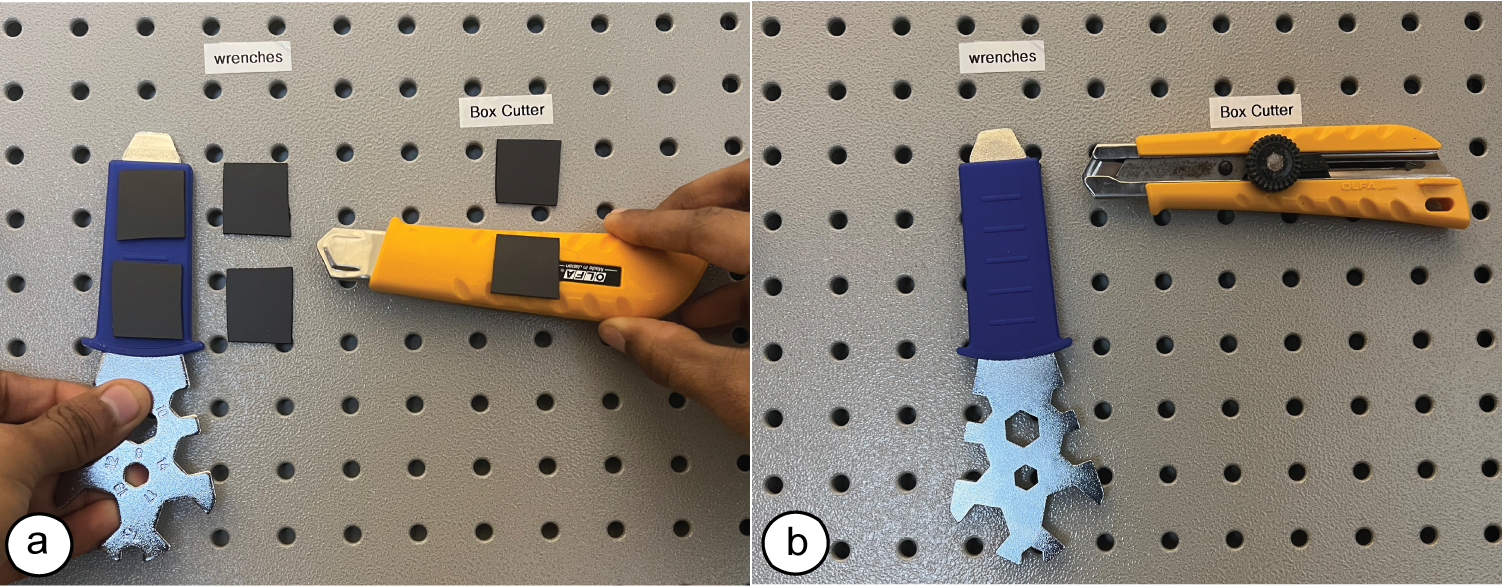}
  \caption{Storage of objects that will only adhere to a workshop wall in particular locations and orientations.}
  \Description{sorting}
  \label{fig:sorting}
\end{figure}

\vspace{5pt}
\noindent{\textbf{Attachments with Specific Orientations:} Figure~\ref{fig:sorting} shows how we can magnetically program attachments that only permit placing objects in specific locations and orientations. We used the user interface's pair generation mode and after plotting the pairs, we attach one magnetic face with double sided tape to a workshop wall and the paired complement to the associated tool. As a result, the tools will only bond when placed at the correct location on the workshop wall and in the correct orientation.}

\begin{figure}[h]
  \centering
  \includegraphics[width=0.9\columnwidth]{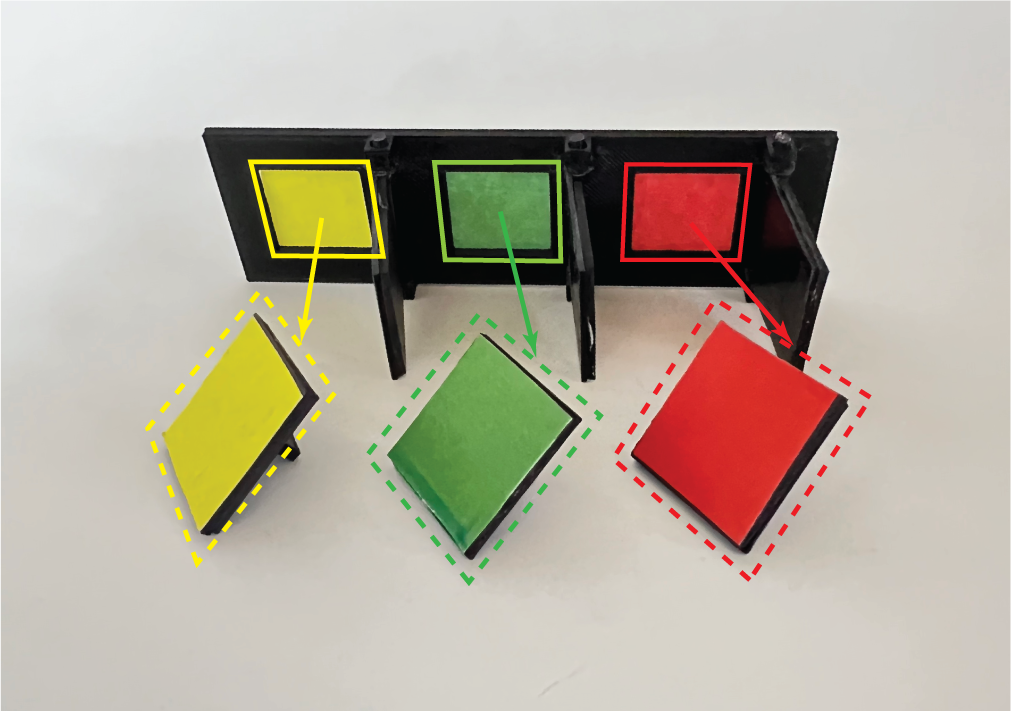}
  \caption{Selectively magnetic lock-and-key systems. Colored locks (solid lines) mate with keys of the associated color (dashed line), but not others.}
  \Description{lock-key}
  \label{fig:doors}
\end{figure}

\vspace{5pt}
\noindent{\textbf{Selectively Pairing Objects:} Figure~\ref{fig:doors} shows how we can use the user interface's pair generation mode to create selectively paired locks and keys. For this, we attach one magnetic face on the key (foreground) and the matching paired complement to the lock (background). For example, the yellow key opens a lock to the hidden yellow surface, but not others. Keys can be programmed to open multiple locks by plotting multiple identical lock patterns, and vice versa. In addition, both locks and keys can be quickly reprogrammed for new security requirements. Moreover, unlike physical keys, the structure of the magnetic keys is invisible to the naked eye. This therefore combines the reprogrammability and imperceptibility of digital keys with the forces of physically actuated keys.}

\begin{figure}[h]
  \centering
  \includegraphics[width=0.9\columnwidth]{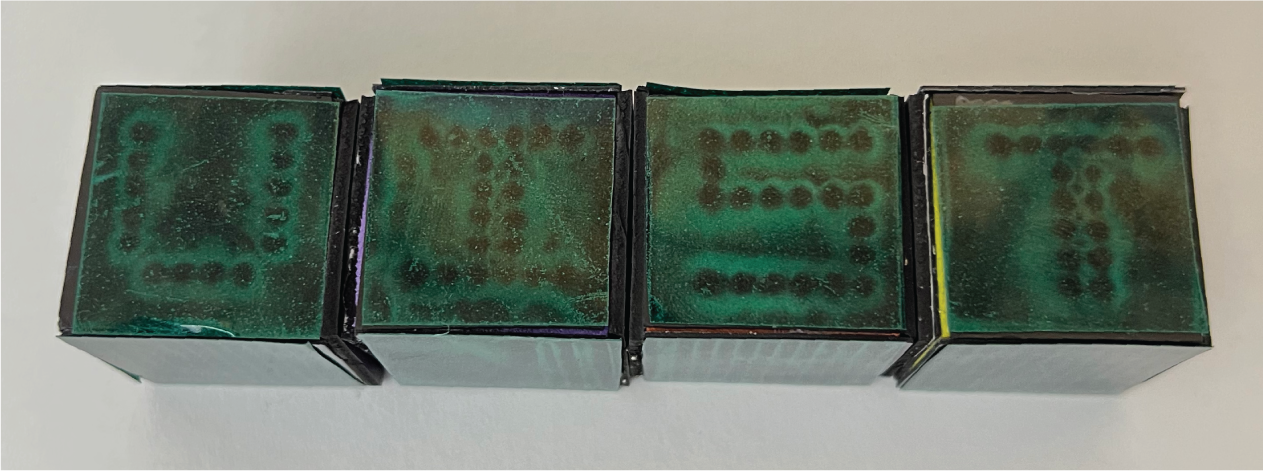}
  \caption{Selectively attractive faces between the cubes guide users into the correct assembly. The letters are plotted onto the top surface of the cubes and made visible with magnetic viewing film.}
  \Description{blocks}
  \label{fig:blocks}
\end{figure}

\vspace{5pt}
\noindent{\textbf{Guided Assembly:} Figure~\ref{fig:blocks} shows how we can create structures that guide the user how to correctly assemble them. In the example shown, four blocks when assembled correctly form the word 'UIST'. To ensure the blocks assemble in the correct way, we use the pair generation mode in the user interface to program selectively mating patterns on the faces \textit{between} cubes. The result is that the cubes adhere magnetically only when assembled correctly to form the word, and cannot be assembled in any other way. To visualize the letters on the top surface, we use the user interface's mode for programming individual pixels to draw the letters and after plotting them onto magnetic sheet material we overlay them with magnetic viewing film to make the magnetic pattern visible to the user. This illustrates the selective patterns' ability to provide affordances that guide human assembly tasks, and our ability to plot arbitrary magnetic patterns as visual textures.}

\begin{figure}[h]
  \centering
  \includegraphics[width=0.9\columnwidth]{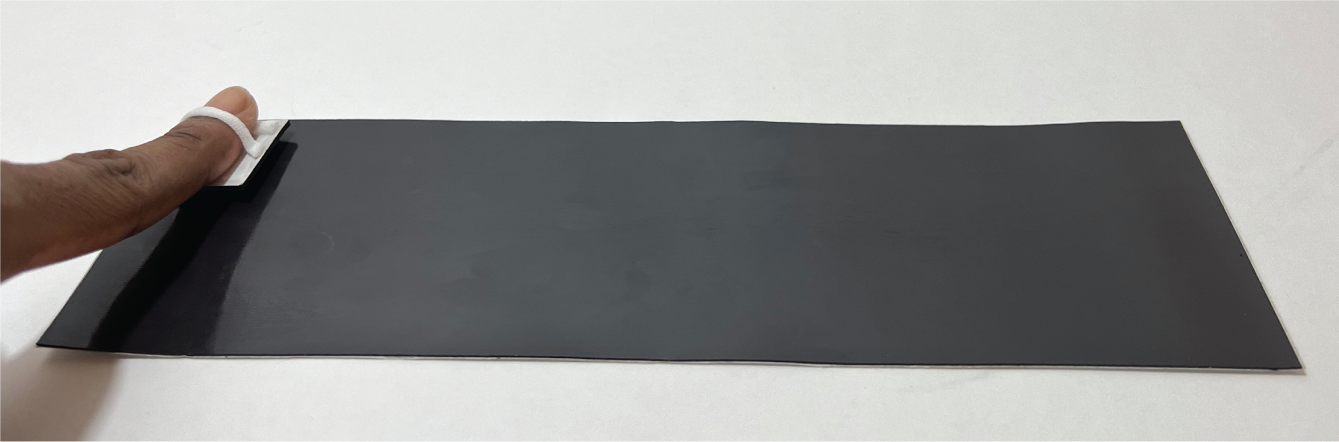}
  \caption{A hand-held token can be programmed to exhibit specific interactions with any location on a programmed sheet.}
  \Description{braille}
  \label{fig:braille}
\end{figure}

\vspace{5pt}
\noindent{\textbf{Haptic Feedback:} Figure~\ref{fig:braille} shows how we can create an arbitrarily large magnetically programmed canvas that provides haptic feedback to a user's finger that has a magnetically programmed token attached to it. The canvas is programmed to exhibit repulsion at chosen locations and agnosticism at others. Coupled with a projector, an image, or mixed reality media, this setup can be utilized to create tactile representations of physical landscapes. } 

\section{Limitations and Future Work}

In the next section, we discuss limitations of our approach and lay out directions for future work.


\vspace{5pt}
\noindent{\textbf{Using the Electromagnet to both Write and Read:}}  We can further simplify the hardware add-on by using the electromagnet to \textit{read} pixels as well as write them. To accomplish this, we can leverage Faraday's law to induce a voltage in the coil as it crosses the pixels' flux lines. This technique has been utilized to both read and write audio tape cassettes from a single electromagnet head for several decades, and thus is promising for our system as well.

\vspace{5pt}
\noindent{\textbf{Continuous Pixels:}} Although our hardware can support pixels with continuous magnetic strength, our user interface and control software currently only support binary pixels of 'North' or 'South' direction. For future work, we plan to extend the user interface by allowing users to assign continuous pixel values by visualizing these in grayscale and then tailoring the current through the electromagnet accordingly during plotting via the control software.

\vspace{5pt}
\noindent{\textbf{Increasing Plotting Speed:}} To accelerate plotting speed, we can affix multiple electromagnets in an array onto the add-on, which will allow us to program multiple pixels in parallel. 

\vspace{5pt}
\noindent{\textbf{Increasing Magnetic Pixel Strength:}} For future work, we will investigate how to increase the magnetic pixel strength by using materials capable of supporting higher magnetic flux densities combined with more powerfully coercive magnetic plotters. This would permit great pixel strengths for the same programming times but may require more complex electronics.

\vspace{5pt}
\noindent{\textbf{Printing 3D Magnetic Structures:}} Recent advances in 3D printing filament have also produced magnetic filament. For future work, we will explore how we can leverage such magnetic filament to both print the structure and as well as the magnetic behavior in one pass to program 3D magnetic objects.


\section{Conclusion}

In this paper, we presented Mixels, programmable magnetic pixels that can be rapidly fabricated using an electromagnetic printhead mounted on an off-the-shelve 3-axis CNC machine. We demonstrated how Mixel's design and fabrication pipeline allows users to quickly create applications with desired magnetic behaviors. We discussed how Mixel's hardware add-on allows to both write and read magnetic pixels and how Mixel's user interface abstracts away the underlying pixel assignment and allows users to specify high-level behaviors, such as selective attraction and repulsion, and otherwise agnostic behavior. We showed in our evaluation that our electromagnet can reliably create pixels of specific magnetic strength on magnetic sheet material, that the pixels exhibit maximum magnetic strength and that the material remains magnetized without attenuation after programming. Finally, we showed a range of applications from selective pairing to guided assembly and haptics. For future work, we plan to further optimize the hardware to use the electromagnet to both write and read magnetic pixel values, to accelerate plotting times and increase magnetic pixel strength, and to explore fabricating 3D geometries with magnetic behaviors in one go by 3D printing with magnetic filament.


\balance
\bibliographystyle{ACM-Reference-Format}
\bibliography{references}


\begin{thebibliography}{46}


\ifx \showCODEN    \undefined \def \showCODEN     #1{\unskip}     \fi
\ifx \showDOI      \undefined \def \showDOI       #1{#1}\fi
\ifx \showISBNx    \undefined \def \showISBNx     #1{\unskip}     \fi
\ifx \showISBNxiii \undefined \def \showISBNxiii  #1{\unskip}     \fi
\ifx \showISSN     \undefined \def \showISSN      #1{\unskip}     \fi
\ifx \showLCCN     \undefined \def \showLCCN      #1{\unskip}     \fi
\ifx \shownote     \undefined \def \shownote      #1{#1}          \fi
\ifx \showarticletitle \undefined \def \showarticletitle #1{#1}   \fi
\ifx \showURL      \undefined \def \showURL       {\relax}        \fi
\providecommand\bibfield[2]{#2}
\providecommand\bibinfo[2]{#2}
\providecommand\natexlab[1]{#1}
\providecommand\showeprint[2][]{arXiv:#2}

\bibitem[\protect\citeauthoryear{??}{pol}{[n.d.]}]%
        {polymagnets}
 \bibinfo{year}{[n.d.]}\natexlab{}.
\newblock \showarticletitle{Polymagnets} \emph{(\bibinfo{series}{last accessed
  July 26, 2022})}.
\newblock
\urldef\tempurl%
\url{http://www.polymagnet.com/polymagnets/}
\showURL{%
\tempurl}


\bibitem[\protect\citeauthoryear{Abe and Sakamoto}{Abe and Sakamoto}{2021}]%
        {magnetrack2021}
\bibfield{author}{\bibinfo{person}{Takehiro Abe} {and} \bibinfo{person}{Daisuke
  Sakamoto}.} \bibinfo{year}{2021}\natexlab{}.
\newblock \bibinfo{booktitle}{\emph{MagneTrack: Magnetic Field Separation
  Method for Continuous and Simultaneous 1-DOF Tracking of Two-Magnets}}.
\newblock \bibinfo{publisher}{Association for Computing Machinery},
  \bibinfo{address}{New York, NY, USA}.
\newblock
\showISBNx{9781450383288}
\urldef\tempurl%
\url{https://doi.org/10.1145/3447526.3472052}
\showURL{%
\tempurl}


\bibitem[\protect\citeauthoryear{Baltieri, Vezzani, and Cucchiara}{Baltieri
  et~al\mbox{.}}{2010}]%
        {baltieri2010fast}
\bibfield{author}{\bibinfo{person}{Davide Baltieri}, \bibinfo{person}{Roberto
  Vezzani}, {and} \bibinfo{person}{Rita Cucchiara}.}
  \bibinfo{year}{2010}\natexlab{}.
\newblock \showarticletitle{Fast background initialization with recursive
  Hadamard transform}. In \bibinfo{booktitle}{\emph{2010 7th IEEE International
  Conference on Advanced Video and Signal Based Surveillance}}. IEEE,
  \bibinfo{pages}{165--171}.
\newblock


\bibitem[\protect\citeauthoryear{Bdeir}{Bdeir}{2009}]%
        {littlebits2009}
\bibfield{author}{\bibinfo{person}{Ayah Bdeir}.}
  \bibinfo{year}{2009}\natexlab{}.
\newblock \showarticletitle{Electronics as Material: LittleBits}. In
  \bibinfo{booktitle}{\emph{Proceedings of the 3rd International Conference on
  Tangible and Embedded Interaction}} (Cambridge, United Kingdom)
  \emph{(\bibinfo{series}{TEI '09})}. \bibinfo{publisher}{Association for
  Computing Machinery}, \bibinfo{address}{New York, NY, USA},
  \bibinfo{pages}{397–400}.
\newblock
\showISBNx{9781605584935}
\urldef\tempurl%
\url{https://doi.org/10.1145/1517664.1517743}
\showDOI{\tempurl}


\bibitem[\protect\citeauthoryear{Bianchi and Oakley}{Bianchi and
  Oakley}{2013}]%
        {magneticappcessories2013}
\bibfield{author}{\bibinfo{person}{Andrea Bianchi} {and} \bibinfo{person}{Ian
  Oakley}.} \bibinfo{year}{2013}\natexlab{}.
\newblock \showarticletitle{Designing Tangible Magnetic Appcessories}. In
  \bibinfo{booktitle}{\emph{Proceedings of the 7th International Conference on
  Tangible, Embedded and Embodied Interaction}} (Barcelona, Spain)
  \emph{(\bibinfo{series}{TEI '13})}. \bibinfo{publisher}{Association for
  Computing Machinery}, \bibinfo{address}{New York, NY, USA},
  \bibinfo{pages}{255–258}.
\newblock
\showISBNx{9781450318983}
\urldef\tempurl%
\url{https://doi.org/10.1145/2460625.2460667}
\showDOI{\tempurl}


\bibitem[\protect\citeauthoryear{Bianchi and Oakley}{Bianchi and
  Oakley}{2015}]%
        {magnid2015}
\bibfield{author}{\bibinfo{person}{Andrea Bianchi} {and} \bibinfo{person}{Ian
  Oakley}.} \bibinfo{year}{2015}\natexlab{}.
\newblock \showarticletitle{MagnID: Tracking Multiple Magnetic Tokens}. In
  \bibinfo{booktitle}{\emph{Proceedings of the Ninth International Conference
  on Tangible, Embedded, and Embodied Interaction}} (Stanford, California, USA)
  \emph{(\bibinfo{series}{TEI '15})}. \bibinfo{publisher}{Association for
  Computing Machinery}, \bibinfo{address}{New York, NY, USA},
  \bibinfo{pages}{61–68}.
\newblock
\showISBNx{9781450333054}
\urldef\tempurl%
\url{https://doi.org/10.1145/2677199.2680582}
\showDOI{\tempurl}


\bibitem[\protect\citeauthoryear{Boldu, Wijewardena, Zhang, and
  Nanayakkara}{Boldu et~al\mbox{.}}{2020}]%
        {maghair2020}
\bibfield{author}{\bibinfo{person}{Roger Boldu}, \bibinfo{person}{Mevan
  Wijewardena}, \bibinfo{person}{Haimo Zhang}, {and} \bibinfo{person}{Suranga
  Nanayakkara}.} \bibinfo{year}{2020}\natexlab{}.
\newblock \bibinfo{booktitle}{\emph{MAGHair: A Wearable System to Create Unique
  Tactile Feedback by Stimulating Only the Body Hair}}.
\newblock \bibinfo{publisher}{Association for Computing Machinery},
  \bibinfo{address}{New York, NY, USA}.
\newblock
\showISBNx{9781450375160}
\urldef\tempurl%
\url{https://doi.org/10.1145/3379503.3403545}
\showURL{%
\tempurl}


\bibitem[\protect\citeauthoryear{Cheung and Girouard}{Cheung and
  Girouard}{2019}]%
        {magneticring2019}
\bibfield{author}{\bibinfo{person}{Victor Cheung} {and} \bibinfo{person}{Audrey
  Girouard}.} \bibinfo{year}{2019}\natexlab{}.
\newblock \showarticletitle{Tangible Around-Device Interaction Using Rotatory
  Gestures with a Magnetic Ring}. In \bibinfo{booktitle}{\emph{Proceedings of
  the 21st International Conference on Human-Computer Interaction with Mobile
  Devices and Services}} (Taipei, Taiwan) \emph{(\bibinfo{series}{MobileHCI
  '19})}. \bibinfo{publisher}{Association for Computing Machinery},
  \bibinfo{address}{New York, NY, USA}, Article \bibinfo{articleno}{26},
  \bibinfo{numpages}{8}~pages.
\newblock
\showISBNx{9781450368254}
\urldef\tempurl%
\url{https://doi.org/10.1145/3338286.3340137}
\showDOI{\tempurl}


\bibitem[\protect\citeauthoryear{Hwang, Ahn, and Wohn}{Hwang
  et~al\mbox{.}}{2013}]%
        {maggetz2013}
\bibfield{author}{\bibinfo{person}{Sungjae Hwang}, \bibinfo{person}{Myungwook
  Ahn}, {and} \bibinfo{person}{Kwang-yun Wohn}.}
  \bibinfo{year}{2013}\natexlab{}.
\newblock \showarticletitle{MagGetz: Customizable Passive Tangible Controllers
  on and around Conventional Mobile Devices}. In
  \bibinfo{booktitle}{\emph{Proceedings of the 26th Annual ACM Symposium on
  User Interface Software and Technology}} (St. Andrews, Scotland, United
  Kingdom) \emph{(\bibinfo{series}{UIST '13})}. \bibinfo{publisher}{Association
  for Computing Machinery}, \bibinfo{address}{New York, NY, USA},
  \bibinfo{pages}{411–416}.
\newblock
\showISBNx{9781450322683}
\urldef\tempurl%
\url{https://doi.org/10.1145/2501988.2501991}
\showDOI{\tempurl}


\bibitem[\protect\citeauthoryear{Ishiguro and Poupyrev}{Ishiguro and
  Poupyrev}{2014}]%
        {printedspeakers2014}
\bibfield{author}{\bibinfo{person}{Yoshio Ishiguro} {and} \bibinfo{person}{Ivan
  Poupyrev}.} \bibinfo{year}{2014}\natexlab{}.
\newblock \showarticletitle{3D Printed Interactive Speakers}. In
  \bibinfo{booktitle}{\emph{Proceedings of the SIGCHI Conference on Human
  Factors in Computing Systems}} (Toronto, Ontario, Canada)
  \emph{(\bibinfo{series}{CHI '14})}. \bibinfo{publisher}{Association for
  Computing Machinery}, \bibinfo{address}{New York, NY, USA},
  \bibinfo{pages}{1733–1742}.
\newblock
\showISBNx{9781450324731}
\urldef\tempurl%
\url{https://doi.org/10.1145/2556288.2557046}
\showDOI{\tempurl}


\bibitem[\protect\citeauthoryear{Kim, Yuk, Zhao, Chester, and Zhao}{Kim
  et~al\mbox{.}}{2018}]%
        {kim2018printing}
\bibfield{author}{\bibinfo{person}{Yoonho Kim}, \bibinfo{person}{Hyunwoo Yuk},
  \bibinfo{person}{Ruike Zhao}, \bibinfo{person}{Shawn~A Chester}, {and}
  \bibinfo{person}{Xuanhe Zhao}.} \bibinfo{year}{2018}\natexlab{}.
\newblock \showarticletitle{Printing ferromagnetic domains for untethered
  fast-transforming soft materials}.
\newblock \bibinfo{journal}{\emph{Nature}} \bibinfo{volume}{558},
  \bibinfo{number}{7709} (\bibinfo{year}{2018}), \bibinfo{pages}{274--279}.
\newblock


\bibitem[\protect\citeauthoryear{Kuo, Liang, Lin, and Chen}{Kuo
  et~al\mbox{.}}{2016}]%
        {gaussmarbles2016}
\bibfield{author}{\bibinfo{person}{Han-Chih Kuo}, \bibinfo{person}{Rong-Hao
  Liang}, \bibinfo{person}{Long-Fei Lin}, {and} \bibinfo{person}{Bing-Yu
  Chen}.} \bibinfo{year}{2016}\natexlab{}.
\newblock \showarticletitle{GaussMarbles: Spherical Magnetic Tangibles for
  Interacting with Portable Physical Constraints}. In
  \bibinfo{booktitle}{\emph{Proceedings of the 2016 CHI Conference on Human
  Factors in Computing Systems}} (San Jose, California, USA)
  \emph{(\bibinfo{series}{CHI '16})}. \bibinfo{publisher}{Association for
  Computing Machinery}, \bibinfo{address}{New York, NY, USA},
  \bibinfo{pages}{4228–4232}.
\newblock
\showISBNx{9781450333627}
\urldef\tempurl%
\url{https://doi.org/10.1145/2858036.2858559}
\showDOI{\tempurl}


\bibitem[\protect\citeauthoryear{Langerak, Z\'{a}rate, Lindlbauer, Holz, and
  Hilliges}{Langerak et~al\mbox{.}}{2020}]%
        {omni2020}
\bibfield{author}{\bibinfo{person}{Thomas Langerak},
  \bibinfo{person}{Juan~Jos\'{e} Z\'{a}rate}, \bibinfo{person}{David
  Lindlbauer}, \bibinfo{person}{Christian Holz}, {and} \bibinfo{person}{Otmar
  Hilliges}.} \bibinfo{year}{2020}\natexlab{}.
\newblock \bibinfo{booktitle}{\emph{Omni: Volumetric Sensing and Actuation of
  Passive Magnetic Tools for Dynamic Haptic Feedback}}.
\newblock \bibinfo{publisher}{Association for Computing Machinery},
  \bibinfo{address}{New York, NY, USA}, \bibinfo{pages}{594–606}.
\newblock
\showISBNx{9781450375146}
\urldef\tempurl%
\url{https://doi.org/10.1145/3379337.3415589}
\showURL{%
\tempurl}


\bibitem[\protect\citeauthoryear{Lee, Post, and Ishii}{Lee
  et~al\mbox{.}}{2011}]%
        {zeron2011}
\bibfield{author}{\bibinfo{person}{Jinha Lee}, \bibinfo{person}{Rehmi Post},
  {and} \bibinfo{person}{Hiroshi Ishii}.} \bibinfo{year}{2011}\natexlab{}.
\newblock \showarticletitle{ZeroN: Mid-Air Tangible Interaction Enabled by
  Computer Controlled Magnetic Levitation}. In
  \bibinfo{booktitle}{\emph{Proceedings of the 24th Annual ACM Symposium on
  User Interface Software and Technology}} (Santa Barbara, California, USA)
  \emph{(\bibinfo{series}{UIST '11})}. \bibinfo{publisher}{Association for
  Computing Machinery}, \bibinfo{address}{New York, NY, USA},
  \bibinfo{pages}{327–336}.
\newblock
\showISBNx{9781450307161}
\urldef\tempurl%
\url{https://doi.org/10.1145/2047196.2047239}
\showDOI{\tempurl}


\bibitem[\protect\citeauthoryear{Leitner and Haller}{Leitner and
  Haller}{2011}]%
        {leitner2011geckos}
\bibfield{author}{\bibinfo{person}{Jakob Leitner} {and}
  \bibinfo{person}{Michael Haller}.} \bibinfo{year}{2011}\natexlab{}.
\newblock \showarticletitle{Geckos: combining magnets and pressure images to
  enable new tangible-object design and interaction}. In
  \bibinfo{booktitle}{\emph{Proceedings of the SIGCHI Conference on Human
  Factors in Computing Systems}}. \bibinfo{pages}{2985--2994}.
\newblock


\bibitem[\protect\citeauthoryear{Liang, Chan, Tseng, Kuo, Huang, Yang, and
  Chen}{Liang et~al\mbox{.}}{2014a}]%
        {gaussbricks2014}
\bibfield{author}{\bibinfo{person}{Rong-Hao Liang}, \bibinfo{person}{Liwei
  Chan}, \bibinfo{person}{Hung-Yu Tseng}, \bibinfo{person}{Han-Chih Kuo},
  \bibinfo{person}{Da-Yuan Huang}, \bibinfo{person}{De-Nian Yang}, {and}
  \bibinfo{person}{Bing-Yu Chen}.} \bibinfo{year}{2014}\natexlab{a}.
\newblock \showarticletitle{GaussBricks: Magnetic Building Blocks for
  Constructive Tangible Interactions on Portable Displays}. In
  \bibinfo{booktitle}{\emph{Proceedings of the SIGCHI Conference on Human
  Factors in Computing Systems}} (Toronto, Ontario, Canada)
  \emph{(\bibinfo{series}{CHI '14})}. \bibinfo{publisher}{Association for
  Computing Machinery}, \bibinfo{address}{New York, NY, USA},
  \bibinfo{pages}{3153–3162}.
\newblock
\showISBNx{9781450324731}
\urldef\tempurl%
\url{https://doi.org/10.1145/2556288.2557105}
\showDOI{\tempurl}


\bibitem[\protect\citeauthoryear{Liang, Cheng, Chan, Peng, Chen, Liang, Yang,
  and Chen}{Liang et~al\mbox{.}}{2013}]%
        {gaussbits2013}
\bibfield{author}{\bibinfo{person}{Rong-Hao Liang}, \bibinfo{person}{Kai-Yin
  Cheng}, \bibinfo{person}{Liwei Chan}, \bibinfo{person}{Chuan-Xhyuan Peng},
  \bibinfo{person}{Mike~Y. Chen}, \bibinfo{person}{Rung-Huei Liang},
  \bibinfo{person}{De-Nian Yang}, {and} \bibinfo{person}{Bing-Yu Chen}.}
  \bibinfo{year}{2013}\natexlab{}.
\newblock \showarticletitle{GaussBits: Magnetic Tangible Bits for Portable and
  Occlusion-Free near-Surface Interactions}. In
  \bibinfo{booktitle}{\emph{Proceedings of the SIGCHI Conference on Human
  Factors in Computing Systems}} (Paris, France) \emph{(\bibinfo{series}{CHI
  '13})}. \bibinfo{publisher}{Association for Computing Machinery},
  \bibinfo{address}{New York, NY, USA}, \bibinfo{pages}{1391–1400}.
\newblock
\showISBNx{9781450318990}
\urldef\tempurl%
\url{https://doi.org/10.1145/2470654.2466185}
\showDOI{\tempurl}


\bibitem[\protect\citeauthoryear{Liang, Kuo, Chan, Yang, and Chen}{Liang
  et~al\mbox{.}}{2014b}]%
        {gaussstones2014}
\bibfield{author}{\bibinfo{person}{Rong-Hao Liang}, \bibinfo{person}{Han-Chih
  Kuo}, \bibinfo{person}{Liwei Chan}, \bibinfo{person}{De-Nian Yang}, {and}
  \bibinfo{person}{Bing-Yu Chen}.} \bibinfo{year}{2014}\natexlab{b}.
\newblock \showarticletitle{GaussStones: Shielded Magnetic Tangibles for
  Multi-Token Interactions on Portable Displays}. In
  \bibinfo{booktitle}{\emph{Proceedings of the 27th Annual ACM Symposium on
  User Interface Software and Technology}} (Honolulu, Hawaii, USA)
  \emph{(\bibinfo{series}{UIST '14})}. \bibinfo{publisher}{Association for
  Computing Machinery}, \bibinfo{address}{New York, NY, USA},
  \bibinfo{pages}{365–372}.
\newblock
\showISBNx{9781450330695}
\urldef\tempurl%
\url{https://doi.org/10.1145/2642918.2647384}
\showDOI{\tempurl}


\bibitem[\protect\citeauthoryear{Mueller, Kruck, and Baudisch}{Mueller
  et~al\mbox{.}}{2013}]%
        {mueller2013laserorigami}
\bibfield{author}{\bibinfo{person}{Stefanie Mueller}, \bibinfo{person}{Bastian
  Kruck}, {and} \bibinfo{person}{Patrick Baudisch}.}
  \bibinfo{year}{2013}\natexlab{}.
\newblock \showarticletitle{LaserOrigami: laser-cutting 3D objects}. In
  \bibinfo{booktitle}{\emph{Proceedings of the SIGCHI Conference on Human
  Factors in Computing Systems}}. \bibinfo{pages}{2585--2592}.
\newblock


\bibitem[\protect\citeauthoryear{Nisser, Cheng, Makaram, Suzuki, and
  Mueller}{Nisser et~al\mbox{.}}{2021a}]%
        {nisser2021programmable}
\bibfield{author}{\bibinfo{person}{Martin Nisser}, \bibinfo{person}{Leon
  Cheng}, \bibinfo{person}{Yashaswini Makaram}, \bibinfo{person}{Ryo Suzuki},
  {and} \bibinfo{person}{Stefanie Mueller}.} \bibinfo{year}{2021}\natexlab{a}.
\newblock \showarticletitle{Programmable Polarities: Actuating Interactive
  Prototypes with Programmable Electromagnets}. In
  \bibinfo{booktitle}{\emph{The Adjunct Publication of the 34th Annual ACM
  Symposium on User Interface Software and Technology}}.
  \bibinfo{pages}{121--123}.
\newblock


\bibitem[\protect\citeauthoryear{Nisser, Cheng, Makaram, Suzuki, and
  Mueller}{Nisser et~al\mbox{.}}{2022a}]%
        {nisser2022electrovoxel}
\bibfield{author}{\bibinfo{person}{Martin Nisser}, \bibinfo{person}{Leon
  Cheng}, \bibinfo{person}{Yashaswini Makaram}, \bibinfo{person}{Ryo Suzuki},
  {and} \bibinfo{person}{Stefanie Mueller}.} \bibinfo{year}{2022}\natexlab{a}.
\newblock \showarticletitle{ElectroVoxel: Electromagnetically Actuated Pivoting
  for Scalable Modular Self-Reconfigurable Robots}. In
  \bibinfo{booktitle}{\emph{2022 International Conference on Robotics and
  Automation (ICRA)}}. IEEE, \bibinfo{pages}{4254--4260}.
\newblock


\bibitem[\protect\citeauthoryear{Nisser, Liao, Chai, Adhikari, Hodges, and
  Mueller}{Nisser et~al\mbox{.}}{2021b}]%
        {nisser2021laserfactory}
\bibfield{author}{\bibinfo{person}{Martin Nisser},
  \bibinfo{person}{Christina~Chen Liao}, \bibinfo{person}{Yuchen Chai},
  \bibinfo{person}{Aradhana Adhikari}, \bibinfo{person}{Steve Hodges}, {and}
  \bibinfo{person}{Stefanie Mueller}.} \bibinfo{year}{2021}\natexlab{b}.
\newblock \showarticletitle{LaserFactory: a laser cutter-based
  electromechanical assembly and fabrication platform to make functional
  devices \& robots}. In \bibinfo{booktitle}{\emph{Proceedings of the 2021 CHI
  Conference on human factors in computing systems}}. \bibinfo{pages}{1--15}.
\newblock


\bibitem[\protect\citeauthoryear{Nisser, Makaram, Faruqi, Suzuki, and
  Mueller}{Nisser et~al\mbox{.}}{2022b}]%
        {Nisser2022preprintIROS}
\bibfield{author}{\bibinfo{person}{Martin Nisser}, \bibinfo{person}{Yashaswini
  Makaram}, \bibinfo{person}{Faraz Faruqi}, \bibinfo{person}{Ryo Suzuki}, {and}
  \bibinfo{person}{Stefanie Mueller}.} \bibinfo{year}{2022}\natexlab{b}.
\newblock \showarticletitle{Selective Self-Assembly using Re-Programmable
  Magnetic Pixels}. In \bibinfo{booktitle}{\emph{2022 IEEE International
  Conference on Intelligent Robots and Systems (IROS)}}. \bibinfo{pages}{To
  Appear}.
\newblock


\bibitem[\protect\citeauthoryear{Nisser, Makaram, and Mueller}{Nisser
  et~al\mbox{.}}{2021c}]%
        {nisser2022stochastic}
\bibfield{author}{\bibinfo{person}{Martin Nisser}, \bibinfo{person}{Yashaswini
  Makaram}, {and} \bibinfo{person}{Stefanie Mueller}.}
  \bibinfo{year}{2021}\natexlab{c}.
\newblock \showarticletitle{Stochastic Self-Assembly with Magnetically
  Re-programmable Voxels}. In \bibinfo{booktitle}{\emph{ACM Symposium on
  Computational Fabrication, Demonstration}}.
\newblock


\bibitem[\protect\citeauthoryear{Ogata}{Ogata}{2018}]%
        {magnetohaptics2018}
\bibfield{author}{\bibinfo{person}{Masa Ogata}.}
  \bibinfo{year}{2018}\natexlab{}.
\newblock \showarticletitle{Magneto-Haptics: Embedding Magnetic Force Feedback
  for Physical Interactions}. In \bibinfo{booktitle}{\emph{Proceedings of the
  31st Annual ACM Symposium on User Interface Software and Technology}}
  (Berlin, Germany) \emph{(\bibinfo{series}{UIST '18})}.
  \bibinfo{publisher}{Association for Computing Machinery},
  \bibinfo{address}{New York, NY, USA}, \bibinfo{pages}{737–743}.
\newblock
\showISBNx{9781450359481}
\urldef\tempurl%
\url{https://doi.org/10.1145/3242587.3242615}
\showDOI{\tempurl}


\bibitem[\protect\citeauthoryear{Ogata and Fukumoto}{Ogata and
  Fukumoto}{2015}]%
        {fluxpaper2015}
\bibfield{author}{\bibinfo{person}{Masa Ogata} {and} \bibinfo{person}{Masaaki
  Fukumoto}.} \bibinfo{year}{2015}\natexlab{}.
\newblock \showarticletitle{FluxPaper: Reinventing Paper with Dynamic Actuation
  Powered by Magnetic Flux}. In \bibinfo{booktitle}{\emph{Proceedings of the
  33rd Annual ACM Conference on Human Factors in Computing Systems}} (Seoul,
  Republic of Korea) \emph{(\bibinfo{series}{CHI '15})}.
  \bibinfo{publisher}{Association for Computing Machinery},
  \bibinfo{address}{New York, NY, USA}, \bibinfo{pages}{29–38}.
\newblock
\showISBNx{9781450331456}
\urldef\tempurl%
\url{https://doi.org/10.1145/2702123.2702516}
\showDOI{\tempurl}


\bibitem[\protect\citeauthoryear{Ogata and Koyama}{Ogata and Koyama}{2021}]%
        {ogata2021computational}
\bibfield{author}{\bibinfo{person}{Masa Ogata} {and} \bibinfo{person}{Yuki
  Koyama}.} \bibinfo{year}{2021}\natexlab{}.
\newblock \showarticletitle{A Computational Approach to Magnetic Force Feedback
  Design}. In \bibinfo{booktitle}{\emph{Proceedings of the 2021 CHI Conference
  on Human Factors in Computing Systems}}. \bibinfo{pages}{1--12}.
\newblock


\bibitem[\protect\citeauthoryear{Pangaro, Maynes-Aminzade, and Ishii}{Pangaro
  et~al\mbox{.}}{2002}]%
        {actuatedworkbench2002}
\bibfield{author}{\bibinfo{person}{Gian Pangaro}, \bibinfo{person}{Dan
  Maynes-Aminzade}, {and} \bibinfo{person}{Hiroshi Ishii}.}
  \bibinfo{year}{2002}\natexlab{}.
\newblock \showarticletitle{The Actuated Workbench: Computer-Controlled
  Actuation in Tabletop Tangible Interfaces}. In
  \bibinfo{booktitle}{\emph{Proceedings of the 15th Annual ACM Symposium on
  User Interface Software and Technology}} (Paris, France)
  \emph{(\bibinfo{series}{UIST '02})}. \bibinfo{publisher}{Association for
  Computing Machinery}, \bibinfo{address}{New York, NY, USA},
  \bibinfo{pages}{181–190}.
\newblock
\showISBNx{1581134886}
\urldef\tempurl%
\url{https://doi.org/10.1145/571985.572011}
\showDOI{\tempurl}


\bibitem[\protect\citeauthoryear{Patten and Ishii}{Patten and Ishii}{2007}]%
        {10.1145/1240624.1240746}
\bibfield{author}{\bibinfo{person}{James Patten} {and} \bibinfo{person}{Hiroshi
  Ishii}.} \bibinfo{year}{2007}\natexlab{}.
\newblock \showarticletitle{Mechanical Constraints as Computational Constraints
  in Tabletop Tangible Interfaces}. In \bibinfo{booktitle}{\emph{Proceedings of
  the SIGCHI Conference on Human Factors in Computing Systems}} (San Jose,
  California, USA) \emph{(\bibinfo{series}{CHI '07})}.
  \bibinfo{publisher}{Association for Computing Machinery},
  \bibinfo{address}{New York, NY, USA}, \bibinfo{pages}{809–818}.
\newblock
\showISBNx{9781595935939}
\urldef\tempurl%
\url{https://doi.org/10.1145/1240624.1240746}
\showDOI{\tempurl}


\bibitem[\protect\citeauthoryear{Pece, Zarate, Vechev, Besse, Gudozhnik, Shea,
  and Hilliges}{Pece et~al\mbox{.}}{2017}]%
        {magtics2017}
\bibfield{author}{\bibinfo{person}{Fabrizio Pece}, \bibinfo{person}{Juan~Jose
  Zarate}, \bibinfo{person}{Velko Vechev}, \bibinfo{person}{Nadine Besse},
  \bibinfo{person}{Olexandr Gudozhnik}, \bibinfo{person}{Herbert Shea}, {and}
  \bibinfo{person}{Otmar Hilliges}.} \bibinfo{year}{2017}\natexlab{}.
\newblock \showarticletitle{MagTics: Flexible and Thin Form Factor Magnetic
  Actuators for Dynamic and Wearable Haptic Feedback}. In
  \bibinfo{booktitle}{\emph{Proceedings of the 30th Annual ACM Symposium on
  User Interface Software and Technology}} (Qu\'{e}bec City, QC, Canada)
  \emph{(\bibinfo{series}{UIST '17})}. \bibinfo{publisher}{Association for
  Computing Machinery}, \bibinfo{address}{New York, NY, USA},
  \bibinfo{pages}{143–154}.
\newblock
\showISBNx{9781450349819}
\urldef\tempurl%
\url{https://doi.org/10.1145/3126594.3126609}
\showDOI{\tempurl}


\bibitem[\protect\citeauthoryear{Peng, Guimbreti\`{e}re, McCann, and
  Hudson}{Peng et~al\mbox{.}}{2016}]%
        {electromagneticprinter2016}
\bibfield{author}{\bibinfo{person}{Huaishu Peng}, \bibinfo{person}{Fran\c{c}ois
  Guimbreti\`{e}re}, \bibinfo{person}{James McCann}, {and}
  \bibinfo{person}{Scott Hudson}.} \bibinfo{year}{2016}\natexlab{}.
\newblock \showarticletitle{A 3D Printer for Interactive Electromagnetic
  Devices}. In \bibinfo{booktitle}{\emph{Proceedings of the 29th Annual
  Symposium on User Interface Software and Technology}} (Tokyo, Japan)
  \emph{(\bibinfo{series}{UIST '16})}. \bibinfo{publisher}{Association for
  Computing Machinery}, \bibinfo{address}{New York, NY, USA},
  \bibinfo{pages}{553–562}.
\newblock
\showISBNx{9781450341899}
\urldef\tempurl%
\url{https://doi.org/10.1145/2984511.2984523}
\showDOI{\tempurl}


\bibitem[\protect\citeauthoryear{Schmitz, Riemann, M\"{u}ller, Kreis, and
  M\"{u}hlh\"{a}user}{Schmitz et~al\mbox{.}}{2021}]%
        {ohsnap2021}
\bibfield{author}{\bibinfo{person}{Martin Schmitz}, \bibinfo{person}{Jan
  Riemann}, \bibinfo{person}{Florian M\"{u}ller}, \bibinfo{person}{Steffen
  Kreis}, {and} \bibinfo{person}{Max M\"{u}hlh\"{a}user}.}
  \bibinfo{year}{2021}\natexlab{}.
\newblock \showarticletitle{Oh, Snap! A Fabrication Pipeline to Magnetically
  Connect Conventional and 3D-Printed Electronics}. In
  \bibinfo{booktitle}{\emph{Proceedings of the 2021 CHI Conference on Human
  Factors in Computing Systems}} (Yokohama, Japan) \emph{(\bibinfo{series}{CHI
  '21})}. \bibinfo{publisher}{Association for Computing Machinery},
  \bibinfo{address}{New York, NY, USA}, Article \bibinfo{articleno}{420},
  \bibinfo{numpages}{11}~pages.
\newblock
\showISBNx{9781450380966}
\urldef\tempurl%
\url{https://doi.org/10.1145/3411764.3445641}
\showDOI{\tempurl}


\bibitem[\protect\citeauthoryear{Strasnick, Yang, Tanner, Olwal, and
  Follmer}{Strasnick et~al\mbox{.}}{2017}]%
        {strasnick2017shiftio}
\bibfield{author}{\bibinfo{person}{Evan Strasnick}, \bibinfo{person}{Jackie
  Yang}, \bibinfo{person}{Kesler Tanner}, \bibinfo{person}{Alex Olwal}, {and}
  \bibinfo{person}{Sean Follmer}.} \bibinfo{year}{2017}\natexlab{}.
\newblock \showarticletitle{shiftio: Reconfigurable tactile elements for
  dynamic affordances and mobile interaction}. In
  \bibinfo{booktitle}{\emph{Proceedings of the 2017 CHI Conference on Human
  Factors in Computing Systems}}. \bibinfo{pages}{5075--5086}.
\newblock


\bibitem[\protect\citeauthoryear{Suzuki, Kato, Gross, and Yeh}{Suzuki
  et~al\mbox{.}}{2018a}]%
        {suzuki2018reactile}
\bibfield{author}{\bibinfo{person}{Ryo Suzuki}, \bibinfo{person}{Jun Kato},
  \bibinfo{person}{Mark~D Gross}, {and} \bibinfo{person}{Tom Yeh}.}
  \bibinfo{year}{2018}\natexlab{a}.
\newblock \showarticletitle{Reactile: Programming swarm user interfaces through
  direct physical manipulation}. In \bibinfo{booktitle}{\emph{Proceedings of
  the 2018 CHI Conference on Human Factors in Computing Systems}}.
  \bibinfo{pages}{1--13}.
\newblock


\bibitem[\protect\citeauthoryear{Suzuki, Stangl, Gross, and Yeh}{Suzuki
  et~al\mbox{.}}{2017}]%
        {suzuki2017fluxmarker}
\bibfield{author}{\bibinfo{person}{Ryo Suzuki}, \bibinfo{person}{Abigale
  Stangl}, \bibinfo{person}{Mark~D Gross}, {and} \bibinfo{person}{Tom Yeh}.}
  \bibinfo{year}{2017}\natexlab{}.
\newblock \showarticletitle{Fluxmarker: Enhancing tactile graphics with dynamic
  tactile markers}. In \bibinfo{booktitle}{\emph{Proceedings of the 19th
  International ACM SIGACCESS Conference on Computers and Accessibility}}.
  \bibinfo{pages}{190--199}.
\newblock


\bibitem[\protect\citeauthoryear{Suzuki, Yamaoka, Leithinger, Yeh, Gross,
  Kawahara, and Kakehi}{Suzuki et~al\mbox{.}}{2018b}]%
        {suzuki2018dynablock}
\bibfield{author}{\bibinfo{person}{Ryo Suzuki}, \bibinfo{person}{Junichi
  Yamaoka}, \bibinfo{person}{Daniel Leithinger}, \bibinfo{person}{Tom Yeh},
  \bibinfo{person}{Mark~D Gross}, \bibinfo{person}{Yoshihiro Kawahara}, {and}
  \bibinfo{person}{Yasuaki Kakehi}.} \bibinfo{year}{2018}\natexlab{b}.
\newblock \showarticletitle{Dynablock: Dynamic 3d printing for instant and
  reconstructable shape formation}. In \bibinfo{booktitle}{\emph{Proceedings of
  the 31st Annual ACM Symposium on User Interface Software and Technology}}.
  \bibinfo{pages}{99--111}.
\newblock


\bibitem[\protect\citeauthoryear{Takahashi, Punpongsanon, and Kim}{Takahashi
  et~al\mbox{.}}{2020}]%
        {programmablefilament2020}
\bibfield{author}{\bibinfo{person}{Haruki Takahashi}, \bibinfo{person}{Parinya
  Punpongsanon}, {and} \bibinfo{person}{Jeeeun Kim}.}
  \bibinfo{year}{2020}\natexlab{}.
\newblock \bibinfo{booktitle}{\emph{Programmable Filament: Printed Filaments
  for Multi-Material 3D Printing}}.
\newblock \bibinfo{publisher}{Association for Computing Machinery},
  \bibinfo{address}{New York, NY, USA}, \bibinfo{pages}{1209–1221}.
\newblock
\showISBNx{9781450375146}
\urldef\tempurl%
\url{https://doi.org/10.1145/3379337.3415863}
\showURL{%
\tempurl}


\bibitem[\protect\citeauthoryear{Torres, Campbell, Kumar, and Paulos}{Torres
  et~al\mbox{.}}{2015}]%
        {hapticprint2015}
\bibfield{author}{\bibinfo{person}{Cesar Torres}, \bibinfo{person}{Tim
  Campbell}, \bibinfo{person}{Neil Kumar}, {and} \bibinfo{person}{Eric
  Paulos}.} \bibinfo{year}{2015}\natexlab{}.
\newblock \showarticletitle{HapticPrint: Designing Feel Aesthetics for Digital
  Fabrication}. In \bibinfo{booktitle}{\emph{Proceedings of the 28th Annual ACM
  Symposium on User Interface Software and Technology}} (Charlotte, NC, USA)
  \emph{(\bibinfo{series}{UIST '15})}. \bibinfo{publisher}{Association for
  Computing Machinery}, \bibinfo{address}{New York, NY, USA},
  \bibinfo{pages}{583–591}.
\newblock
\showISBNx{9781450337793}
\urldef\tempurl%
\url{https://doi.org/10.1145/2807442.2807492}
\showDOI{\tempurl}


\bibitem[\protect\citeauthoryear{Yamaoka and Kakehi}{Yamaoka and
  Kakehi}{2013}]%
        {depend2013}
\bibfield{author}{\bibinfo{person}{Junichi Yamaoka} {and}
  \bibinfo{person}{Yasuaki Kakehi}.} \bibinfo{year}{2013}\natexlab{}.
\newblock \showarticletitle{DePENd: Augmented Handwriting System Using
  Ferromagnetism of a Ballpoint Pen}. In \bibinfo{booktitle}{\emph{Proceedings
  of the 26th Annual ACM Symposium on User Interface Software and Technology}}
  (St. Andrews, Scotland, United Kingdom) \emph{(\bibinfo{series}{UIST '13})}.
  \bibinfo{publisher}{Association for Computing Machinery},
  \bibinfo{address}{New York, NY, USA}, \bibinfo{pages}{203–210}.
\newblock
\showISBNx{9781450322683}
\urldef\tempurl%
\url{https://doi.org/10.1145/2501988.2502017}
\showDOI{\tempurl}


\bibitem[\protect\citeauthoryear{Yasu}{Yasu}{2017}]%
        {yasu2017magnetic}
\bibfield{author}{\bibinfo{person}{Kentaro Yasu}.}
  \bibinfo{year}{2017}\natexlab{}.
\newblock \showarticletitle{Magnetic plotter: a macrotexture design method
  using magnetic rubber sheets}. In \bibinfo{booktitle}{\emph{Proceedings of
  the 2017 CHI Conference on Human Factors in Computing Systems}}.
  \bibinfo{pages}{4983--4993}.
\newblock


\bibitem[\protect\citeauthoryear{Yasu}{Yasu}{2019}]%
        {magnetact2019}
\bibfield{author}{\bibinfo{person}{Kentaro Yasu}.}
  \bibinfo{year}{2019}\natexlab{}.
\newblock \showarticletitle{Magnetact: Magnetic-Sheet-Based Haptic Interfaces
  for Touch Devices}. In \bibinfo{booktitle}{\emph{Proceedings of the 2019 CHI
  Conference on Human Factors in Computing Systems}} (Glasgow, Scotland Uk)
  \emph{(\bibinfo{series}{CHI '19})}. \bibinfo{publisher}{Association for
  Computing Machinery}, \bibinfo{address}{New York, NY, USA},
  \bibinfo{pages}{1–8}.
\newblock
\showISBNx{9781450359702}
\urldef\tempurl%
\url{https://doi.org/10.1145/3290605.3300470}
\showDOI{\tempurl}


\bibitem[\protect\citeauthoryear{Yasu}{Yasu}{2020}]%
        {yasu2020magnelayer}
\bibfield{author}{\bibinfo{person}{Kentaro Yasu}.}
  \bibinfo{year}{2020}\natexlab{}.
\newblock \showarticletitle{MagneLayer: Force Field Fabrication by Layered
  Magnetic Sheets}. In \bibinfo{booktitle}{\emph{Proceedings of the 2020 CHI
  Conference on Human Factors in Computing Systems}}. \bibinfo{pages}{1--9}.
\newblock


\bibitem[\protect\citeauthoryear{Yasu and Ishikawa}{Yasu and Ishikawa}{2021}]%
        {magnetactanimals2021}
\bibfield{author}{\bibinfo{person}{Kentaro Yasu} {and} \bibinfo{person}{Masaya
  Ishikawa}.} \bibinfo{year}{2021}\natexlab{}.
\newblock \bibinfo{booktitle}{\emph{Magnetact Animals: A Simple Kinetic Toy Kit
  for a Creative Online Workshop for Children}}.
\newblock \bibinfo{publisher}{Association for Computing Machinery},
  \bibinfo{address}{New York, NY, USA}.
\newblock
\showISBNx{9781450380959}
\urldef\tempurl%
\url{https://doi.org/10.1145/3411763.3451533}
\showURL{%
\tempurl}


\bibitem[\protect\citeauthoryear{Yasu and Katsumoto}{Yasu and
  Katsumoto}{2015}]%
        {bumpahead2015}
\bibfield{author}{\bibinfo{person}{Kentaro Yasu} {and}
  \bibinfo{person}{Yuichiro Katsumoto}.} \bibinfo{year}{2015}\natexlab{}.
\newblock \showarticletitle{Bump Ahead: Easy-to-Design Haptic Surface Using
  Magnet Array}. In \bibinfo{booktitle}{\emph{SIGGRAPH Asia 2015 Emerging
  Technologies}} (Kobe, Japan) \emph{(\bibinfo{series}{SA '15})}.
  \bibinfo{publisher}{Association for Computing Machinery},
  \bibinfo{address}{New York, NY, USA}, Article \bibinfo{articleno}{3},
  \bibinfo{numpages}{3}~pages.
\newblock
\showISBNx{9781450339254}
\urldef\tempurl%
\url{https://doi.org/10.1145/2818466.2818478}
\showDOI{\tempurl}


\bibitem[\protect\citeauthoryear{Zeng, Deng, Zhu, Wessely, Kilian, and
  Mueller}{Zeng et~al\mbox{.}}{2021}]%
        {lenticular2021}
\bibfield{author}{\bibinfo{person}{Jiani Zeng}, \bibinfo{person}{Honghao Deng},
  \bibinfo{person}{Yunyi Zhu}, \bibinfo{person}{Michael Wessely},
  \bibinfo{person}{Axel Kilian}, {and} \bibinfo{person}{Stefanie Mueller}.}
  \bibinfo{year}{2021}\natexlab{}.
\newblock \showarticletitle{Lenticular Objects: 3D Printed Objects with
  Lenticular Lens Surfaces That Can Change Their Appearance Depending on the
  Viewpoint}. In \bibinfo{booktitle}{\emph{The 34th Annual ACM Symposium on
  User Interface Software and Technology}} (Virtual Event, USA)
  \emph{(\bibinfo{series}{UIST '21})}. \bibinfo{publisher}{Association for
  Computing Machinery}, \bibinfo{address}{New York, NY, USA},
  \bibinfo{pages}{1184–1196}.
\newblock
\showISBNx{9781450386357}
\urldef\tempurl%
\url{https://doi.org/10.1145/3472749.3474815}
\showDOI{\tempurl}


\bibitem[\protect\citeauthoryear{Zheng, Kim, Leithinger, Gross, and Do}{Zheng
  et~al\mbox{.}}{2019}]%
        {zheng2019mechamagnets}
\bibfield{author}{\bibinfo{person}{Clement Zheng}, \bibinfo{person}{Jeeeun
  Kim}, \bibinfo{person}{Daniel Leithinger}, \bibinfo{person}{Mark~D Gross},
  {and} \bibinfo{person}{Ellen Yi-Luen Do}.} \bibinfo{year}{2019}\natexlab{}.
\newblock \showarticletitle{Mechamagnets: Designing and fabricating haptic and
  functional physical inputs with embedded magnets}. In
  \bibinfo{booktitle}{\emph{Proceedings of the Thirteenth International
  Conference on Tangible, Embedded, and Embodied Interaction}}.
  \bibinfo{pages}{325--334}.
\newblock


\end{thebibliography}

\end{document}